\documentclass[journal=jacsat,manuscript=article]{achemso}

\usepackage[version=3]{mhchem} 
\usepackage{xcolor}
\usepackage{adjustbox}
\usepackage{graphics}

\title{Systematic DFT+U and Quantum Monte Carlo benchmark of magnetic two-dimensional (2D) CrX$_3$ (X = I, Br, Cl, F)  }
\author{Daniel Wines}
\affiliation{Materials Science and Engineering
Division, National Institute of Standards and Technology (NIST),
Gaithersburg, MD 20899, USA}
\email{daniel.wines@nist.gov}

\author{Kamal Choudhary}
\affiliation{Materials Science and Engineering
Division, National Institute of Standards and Technology (NIST),
Gaithersburg, MD 20899, USA}
\alsoaffiliation{Theiss Research,
La Jolla, CA, 92037, USA}

\author{Francesca Tavazza}
\affiliation{Materials Science and Engineering
Division, National Institute of Standards and Technology (NIST),
Gaithersburg, MD 20899, USA}

\date{\today}% It is always \today, today,
\begin{document}

\begin{abstract}

The search for two-dimensional (2D) magnetic
materials has attracted a great deal of attention because of the experimental synthesis of 2D CrI$_3$, which has a measured Curie temperature of 45 K. Often times, these monolayers have a higher degree of electron correlation and require more sophisticated methods beyond density functional theory (DFT). Diffusion Monte Carlo (DMC) is a correlated electronic structure method that has been demonstrated successful for calculating the electronic and magnetic properties of a wide variety of 2D and bulk systems, since it has a weaker dependence
on the Hubbard parameter (U) and density functional. In this study we designed a workflow that combines DFT+U and DMC in order to treat 2D correlated magnetic systems. We chose monolayer CrX$_3$ (X = I, Br, Cl, F), with a stronger focus on CrI$_3$ and CrBr$_3$, as a case study due to the fact that they have been experimentally realized and have a finite critical temperature. With this DFT+U and DMC workflow and the analytical method of Torelli and Olsen, we estimated a maximum value of 43.56 K for the T$_c$ of CrI$_3$ and 20.78 K for the T$_c$ of CrBr$_3$, in addition to analyzing the spin densities and magnetic properties with DMC and DFT+U. We expect that running this workflow for a well-known material class will aid in the future discovery and characterization of lesser known and more complex correlated 2D magnetic materials.

\end{abstract}

\maketitle

%\tableofcontents

\section{\label{sec:intro}Introduction}

Recently, the search for two-dimensional (2D) magnetic materials, especially ferromagnets, has become an important task for the materials science community. The revolutionary experimental synthesis of 2D CrI$_3$, which has a measured Curie temperature of 45 K \cite{cri3}, has sparked interest in discovering and utilizing similar ferromagnetic materials for next generation devices. Beyond monolayer CrI$_3$, room temperature magnetism has been experimentally measured for 2D VSe$_2$ on a van der Waals substrate \cite{vse2} and it has been shown that ferromagnetic order exists in 2D Cr$_2$Ge$_2$Te$_6$ \cite{crgete} and Fe$_3$GeTe$_2$ \cite{fe3gete2}. In addition, theoretical calculations have predicted ferromagnetic ordering in monolayers such as CrBr$_3$ \cite{olsen-data}, CrCl$_3$ \cite{olsen-data,PhysRevLett.127.037204}, CrF$_3$ \cite{C5TC02840J}, MnO$_2$ \cite{fm-mno2,mno2-qmc}, FeCl$_2$ \cite{https://doi.org/10.48550/arxiv.2205.00300}, K$_2$CuF$_4$ \cite{PhysRevB.88.201402}, the family of MPX$_3$ (M is 3d transition metal atom, X is group VI atom) \cite{PhysRevB.94.184428}, $\alpha$-RuCl$_3$ \cite{C7CP07953B}, RuBr$_3$ and RuI$_3$ \cite{ERSAN2019111} and several other reported materials \cite{olsen-data}. 

Often times, however, these monolayers have a higher degree of electron correlation and require more sophisticated methods beyond density functional theory (DFT). 
Diffusion Monte Carlo (DMC) \cite{RevModPhys.73.33} is a many-body correlated electronic structure method that has successful for the calculation of electronic and magnetic properties of a variety of bulk and low-dimensional systems \cite{ataca_qmc,PhysRevB.95.081301,PhysRevB.96.075431,PhysRevMaterials.3.124414,Luo_2016,C6CP02067D,doi:10.1063/1.4919242,PhysRevB.98.155130,doi:10.1063/5.0022814,bennett2021origin,PhysRevB.103.205206,PhysRevX.4.031003,PhysRevB.94.035108,PhysRevMaterials.2.085801,PhysRevMaterials.5.024002,doi:10.1063/5.0023223,wines2021pathway,bilayer-phos,PhysRevX.9.011018,PhysRevLett.115.115501,doi:10.1063/1.5026120,phosphors,PhysRevMaterials.1.065408,staros,https://doi.org/10.48550/arxiv.2203.15949}. This method has a much weaker dependence on the starting density functional and Hubbard parameter, can achieve results with an accuracy beyond DFT \cite{RevModPhys.73.33}, and scales similary to DFT with respect to the number of electrons in the simulation \cite{RevModPhys.73.33}. For example, DMC has successfully predicted the magnetic structure for FeSe when DFT methods contradicted \cite{PhysRevB.94.035108}. In addition, the correct spin superexchange in the correlated cuprate Ca$_2$CuO$_3$ has been determined \cite{PhysRevX.4.031003} using DMC methods \cite{PhysRevX.4.031003}. With regards to 2D materials and DMC, the band gap of GaSe \cite{doi:10.1063/5.0023223} has been calculated to be in excellent agreement with experiment, the correct atomic structure and potential energy surface of CrI$_3$ \cite{staros} and GeSe \cite{PhysRevMaterials.5.024002} has been predicted, and the critical temperature of MnO$_2$ \cite{mno2-qmc} has been estimated. 

The last step in the investigation of 2D correlated magnetic systems is the estimation of T$_c$. Specifically, 
the Mermin-Wagner theorem \cite{PhysRevLett.17.1133} implies that magnetic order in a monolayer cannot exist unless magnetic anisotropy (MA) is present and perpendicular to the plane, which allows a finite critical temperature (T$_c$). In order to obtain an appropriate value for T$_c$, the magnetic anisotropy energies (MAE) should be determined by performing  non-collinear (spin-orbit) calculations. Once the MAE and magnetic exchange parameters of a 2D system are obtained from first principles, they can be input into analytical models such as the one derived by Torelli and Olsen \cite{Torelli_2018} to estimate T$_c$. 

Data-driven high throughput studies of 2D materials are needed to identify candidates that meet the criteria to be a 2D ferromagnet with a finite T$_c$. Development of an efficient framework dealing with all the computational steps highlighted above is necessary to make such a high throughput search possible.   

In this work we developed one such workflow and used it to perform a systematic DFT+U and Quantum Monte Carlo benchmark of magnetic two-dimensional (2D). The framework and all data it produced have been made available to the public through the JARVIS (Joint Automated Repository for Various Integrated Simulations, \url{https://jarvis.nist.gov/}) project. JARVIS, which is part of Materials Genome Initiative (MGI), is a computational materials science framework developed at National Institute of
Standards and Technology (NIST) \cite{jarvis}. One of the main components, JARVIS-DFT, is a comprehensive database of DFT-computed material properties for over 3,000 2D and 2D-like materials and over 60,000 bulk materials, with results from multiple functionals such as PBE and vdW-DF-OptB88. JARVIS-DFT contains DFT calculated structural, energetics \cite{PhysRevB.98.014107}, elastic \cite{jarvis-opt}, optoelectronic \cite{jarvis-opt}, thermoelectric \cite{Choudhary_2020}, piezoelectric, dielectric, infrared \cite{jarvis-dfpt}, solar-efficiency \cite{jarvis-solar}, topological \cite{jarvis-topological}, anomalous quantum confinement \cite{PhysRevMaterials.5.054602}, and superconducting \cite{https://doi.org/10.48550/arxiv.2205.00060} properties. As a response to the limitations of DFT, a limited number of beyond-DFT data, such as hybrid functionals (HSE06, PBE0) or many-body (GW, DMFT) results have been added to JARVIS, to increase accuracy and reliability of data. The discrepancies observed in DFT results, between different density functionals and experimental data, are more prevalent for materials that have a higher degree of electron correlation, and are common for the 2D magnetic structures of interest. Specifically, this failure of DFT to describe correlated systems can be due in part to the tendency of standard exchange-correlation functionals to over-delocalize valence electrons \cite{https://doi.org/10.1002/qua.24521}. That is why DFT fails for systems whose ground state is characterized by a more pronounced localization of electrons, such as transition metal-based materials. This delocalization occurs due to the inability of the exchange-correlation functional to completely cancel out the electronic self-interaction, and a ``fragment” of the same electron remains that can induce added self-interaction, inducing an excessive delocalization of the wavefunction \cite{https://doi.org/10.1002/qua.24521}. To combat this self-interaction error, more sophisticated density functionals such as meta-GGAs \cite{PhysRevLett.115.036402,r2scan}, hybrid functionals \cite{doi:10.1063/1.1564060}, or DFT functionals with the added Hubbard (U) correction \cite{PhysRevB.57.1505} can be utilized. Despite the fact that these more sophisticated DFT methods exist, often times the calculated properties of 2D magnets are heavily influenced by which density functional and U parameter are used. Due to this, a method that has a weaker dependence on the U parameter and functional and can capture the electron correlation that drive magnetic ordering is required. With such a many-body method, the realization of 2D magnetic device fabrication can be significantly expedited.

\begin{figure}
\begin{center}
\includegraphics[width=8cm]{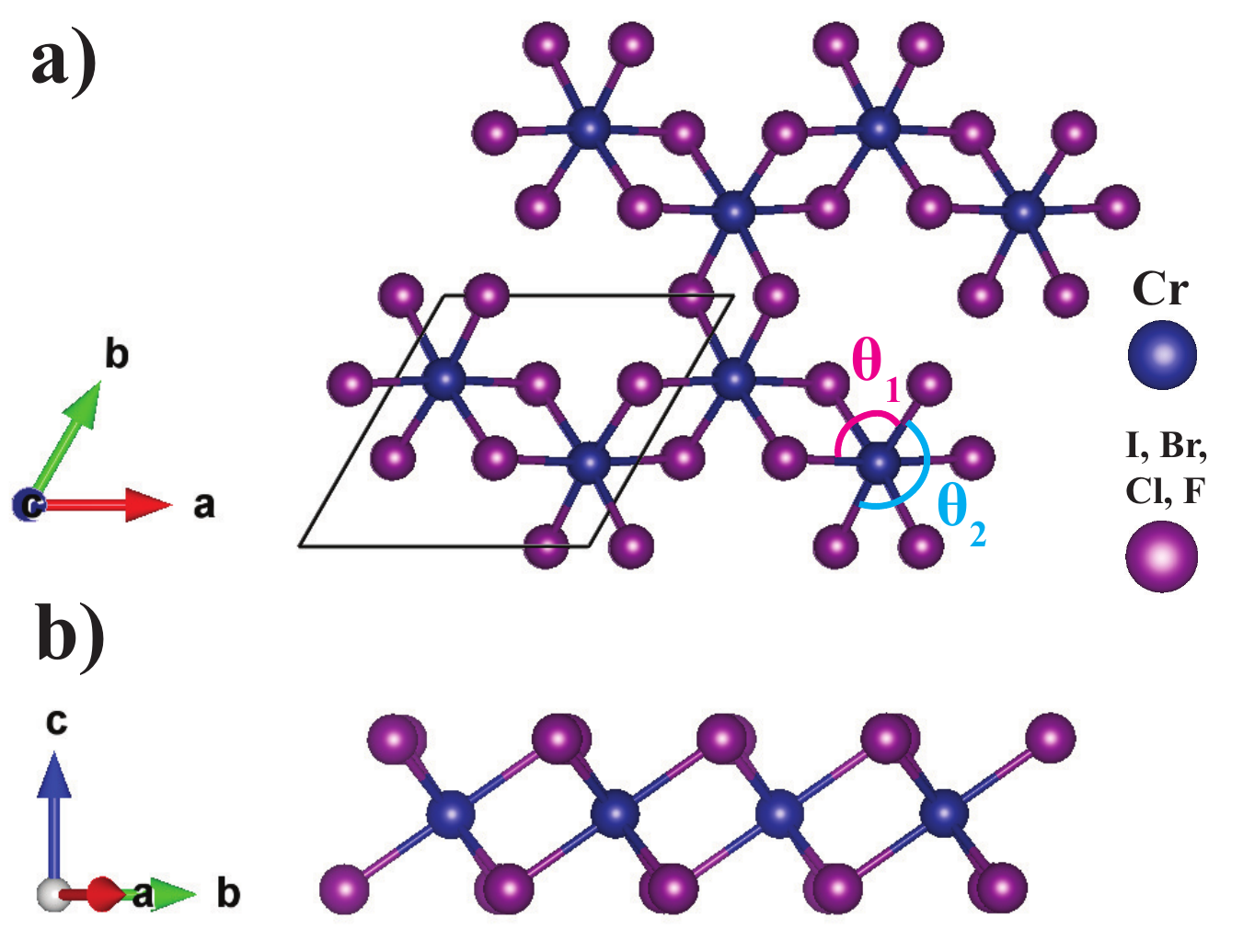}
\caption{Top a) and side b) view of the atomic structure of monolayer CrX$_3$ (X = I, Br, Cl, F). More detailed structural parameters are given in Table S2. }
\label{structure}
\end{center}
\end{figure}

The goal of this study is to utilize a higher order many-body method such as DMC, in order to treat 2D correlated magnetic systems. We chose monolayer CrX$_3$ (X = I, Br, Cl, F) as a case study because they are an ideal class of materials to benchmark, due to the fact that they have been experimentally realized \cite{cri3,crbr-exp,crcl3-exp}, have a finite critical temperature \cite{cri3,olsen-data}, and have extensively been studied with DFT \cite{olsen-data}. We hope that making such a DMC framework public through the JARVIS project, for a well-known class of materials will aid in the future discovery and characterization of lesser known and more complex correlated 2D materials. 

The outline of this paper is as follows: The Computational Methods section will outline the computational approaches for DFT, Quantum Monte Carlo (QMC), post-processing of QMC data, and will detail the newly developed DFT-QMC JARVIS workflow to obtain accurate magnetic properties of 2D systems. The Results and Discussion section will present DFT benchmarking using various DFT methods, and QMC results, and the Conclusion section will provide concluding remarks.

\section{Computational Methods}\label{methods}

\subsection{\label{sec:dft-qmcmethods}DFT and QMC Methods}

We performed DFT calculations with the added Hubbard correction (U) \cite{PhysRevB.57.1505} to treat the on-site Coulomb interaction of the $3d$ orbitals of Cr atoms. To determine how the choice of functional impacts the results, benchmarking DFT simulations were performed using the Vienna Ab initio Simulation (VASP) code and projector augmented wave (PAW) pseudopotentials \cite{PhysRevB.54.11169,PhysRevB.59.1758} (please note that the use of commercial software (VASP) does not imply recommendation by the National Institute of Standards and Technology). It is advantageous to benchmark these materials with VASP and PAW pseudopotentials due to the fact that they require a much smaller cutoff energy and are therefore more cost effective for a large number of simulations. For these reference calculations, the Perdew-Burke-Ernzerhof (PBE)\cite{PhysRevLett.77.3865}, local density approximation (LDA)\cite{PhysRev.136.B864} the strongly constrained and appropriately normed (SCAN)\cite{PhysRevLett.115.036402} meta-GGA, and recently developed r2SCAN \cite{r2scan} functionals were used. r2SCAN was developed to improve the numerical performance of SCAN, at the expense of breaking constraints known from the exact exchange–correlation functional \cite{r2scan}. In addition, to increase accuracy, we also performed calculations with the screened hybrid HSE06 functional, which is created by mixing 75 $\%$ of the PBE exchange with 25 $\%$ of the Fock
exchange and 100 $\%$ of the correlation energy from PBE \cite{doi:10.1063/1.1564060}. For our VASP calculations using PBE, LDA, SCAN, and r2SCAN (+U), we used the Cr$_{pv}$ PAW potential (12 electrons) while for HSE06, we used the standard Cr PAW potential (6 electrons) due to computational restraints. We justified this choice for the HSE06 calculations by performing PBE+U (U = 2 eV) calculations with the Cr$_{pv}$ PAW potential and the Cr standard PAW potential separately, and found the results for magnetic exchange to be within 2-4 $\%$ of each other. There was at least 20 \AA\space of vacuum given between periodic layers of CrX$_3$ in the $c$-direction. We used a kinetic energy cutoff of 500 eV for CrI$_3$, CrBr$_3$ and CrCl$_3$ and a cutoff of 700 eV for CrF$_3$. A 5x5x1 k-point grid was used for the 8 atom unit cell of all CrX$_3$ monolayers. To calculate MAE, spin-orbit DFT (DFT+U) calculations were carried out for the FM and AFM states of each 2D CrX$_3$. This is done by performing two spin-orbit calculations, one where the spins are oriented in the off-plane direction (in our case, $z$) and one where the easy axis is rotated 90$^{\circ}$ (in our case, $x$). A workflow where these four static spin-orbit calculations (for the FM and AFM phases) are performed (using the previously optimized vdW-DF-OptB88 \cite{Klime__2009} geometry) was carried out within JARVIS.

For all QMC calculations, we used DFT and the PBE functional to create the trial wavefunction for subsequent fixed-node DMC calculations. The Quantum Espresso (QE) \cite{Giannozzi_2009} code was used for the DFT calculations within our QMC workflow. The trial wavefunction was generated separately for the FM and AFM configurations of each CrX$_3$ monolayer, using various values of U. The reason for this was to variationally determine the optimal nodal surface (find the value of U that produces the lowest total energy). All QMC calculations require norm-conserving (NC) pseudopotentials. For Cr, we used hard NC RRKJ (OPT), scalar relativistic pseudopotentials \cite{PhysRevB.93.075143}. For Br, Cl and F, scalar relativistic effective core potentials (ECP) were used \cite{doi:10.1063/1.5038135,doi:10.1063/1.5121006}. For I, a newly developed averaged relativistic effective potential (AREP) was used \cite{https://doi.org/10.48550/arxiv.2202.04747}. For DMC calculations that explicitly include spin-orbit effects, a spin-orbit relativistic effective potential (SOREP) for I \cite{https://doi.org/10.48550/arxiv.2202.04747} could be used, but since the spin-orbit interaction of CrI$_3$ are on the order of the DMC error bar, we decided to use the AREP potential for I. These NC potentials are meant to produce all electron results (in accordance with coupled cluster theory), while the PAW pseudopotentials are meant to reproduce all electron density functional results. Ideally, the energy differences obtained with either method should yield similar results, which we verify by observing a 2 - 6 $\%$ difference between PAW results and NC results for magnetic exchange (for PBE+U, U = 2 eV). For these pseudopotentials, we used a kinetic energy cutoff of 300 Ry ($\approx$ 4,080 eV) for all calculations, with the exception of F which required 600 Ry ($\approx$ 8,160 eV, see Fig. S1). We tested the reciprocal grid size at the DFT level and determined that a k-grid of 3x3x1 was sufficient for each CrX$_3$ monolayer (see Fig. S2). Although three different types of pseudopotentials were used in our calculations (PAW for DFT benchmarking, PAW$_{pv}$ for HSE06 benchmarking, and NC for DMC), the results at the DFT level are all within 2 - 8 $\%$ of each other, which indicates that pseudopotential choice does not significantly hinder the accuracy of such small energy scale calculations.   

The QMCPACK \cite{Kim_2018,doi:10.1063/5.0004860} code was used to carry out Variational Monte Carlo (VMC) and DMC \cite{RevModPhys.73.33,Needs_2009} calculations, after the DFT generation of the trial wavefunction. VMC calculations are the intermediate steps between the DFT and DMC calculations, where the single determinant wavefunction from DFT is converted into a many-body wavefunction by use of the Jastrow parameters \cite{PhysRev.34.1293,PhysRev.98.1479}, which aid in modeling the electron correlation and ultimately reduce the uncertainty in the DMC simulations \cite{PhysRevLett.94.150201,doi:10.1063/1.460849}. Up to two-body Jastrow \cite{PhysRevB.70.235119} correlation functions were included in the trial wavefunction. The linear method \cite{PhysRevLett.98.110201} was used to minimize the variance and energy respectively of the VMC energies. The cost function of the variance optimization is 100 $\%$ variance minimization while the cost function of the energy optimization is split as 95 $\%$ energy minimization and 5 $\%$ variance minimization, which has been demonstrated to reduce the uncertainty for DMC results \cite{PhysRevLett.94.150201}. The DFT-VMC-DMC workflows were automated using the Nexus \cite{nexus} software suite. After testing, a large supercell size of 48 atoms was deemed to be sufficient in eliminating finite-size effects for the FM and AFM configurations of 2D CrI$_3$ (see Fig. S3), which justified our choice to use a 48 atom supercell to calculate the magnetic exchange energy with QMC for the other materials of interest. For the DMC simulations, the T-moves \cite{doi:10.1063/1.3380831} algorithm was used to evaluate the nonlocal part of the pseudopotentials and after testing, a timestep of 0.01 Ha$^{-1}$ was determined to be sufficient (see Table S1). The error in our DMC calculations is the standard error about the mean value. This $\pm$ value about the mean is indicated by the error bars in the figures and the parenthesis in the tabulated results. As previously stated, in our DMC simulations, controllable errors such as timestep, finite-size, and choice of nodal surface are appropriately addressed. The remaining sources of uncertainty can arise from the fixed-node bias and the error associated with evaluating the nonlocal part of the pseudopotential. Luckily, the fixed-node error has been demonstrated to be on the order of 1 - 2 $\%$ in other crystal systems \cite{PhysRevB.103.205206} and the T-moves \cite{doi:10.1063/1.3380831} algorithm has been proved to be successful in evaluating the nonlocal part of the pseudopotential, since it is treated in a variational way \cite{doi:10.1063/1.3380831,PhysRevB.103.205206}. 

The total charge and spin densities were extracted from our DMC results. The spin density ($\rho_s$) is the difference between the spin-up contribution to the total charge density and the spin-down contribution to the total charge density ($\rho_s = \rho_{up}-\rho_{down}$). We used an extrapolation scheme on the DMC densities in order to eliminate the bias that occurs from using a mixed estimator. Because the density estimator does not commute with the fixed-node Hamiltonian, the DMC density we calculated is a mixed estimator between the pure fixed-node DMC and VMC densities. The extrapolation formula takes the form \cite{RevModPhys.73.33}:

\begin{equation} \label{rho1}
\rho_1 =2\rho_{\textrm{DMC}}-\rho_{\textrm{VMC}}+\mathcal{O}[(\Phi-\Psi_{\textrm{T}})^2]
\end{equation}
where $\rho_{\textrm{DMC}}$ and $\rho_{\textrm{VMC}}$ are the DMC and VMC charge densities respectively. $\Phi$ is the trial wavefunction arising from the DMC Hamiltonian and $\Psi_{\textrm{T}}$ is the trial wavefunction arising from VMC. 

We integrated the DFT (DFT+U) and DMC spin densities up to a cutoff radius $r_{cut}$ (which is defined as half of the Cr-X bond distance in CrX$_3$ monolayers) to obtain an estimate for the site-averaged atomic magnetic moment per atom. To calculate these magnetic moments per atom ($M_A$), we summed over the spherically interpolated spin densities:

\begin{equation} \label{M_a}
M_A = 4\pi \int_0^{r_{cut}}r^2 \rho_s(r)dr \approx 4\pi \sum_{i=0}^{r_{cut}/\Delta r}r_i^2 \rho_s(r_i)\Delta r
\end{equation}
where $\Delta r$ is the radial grid size and $r_i$ is the distance from the center of the atom to a given point on the grid.

To estimate the critical temperature of CrX$_3$ monolayers, we used the method outlined by Torelli and Olsen \cite{Torelli_2018}, which derived a simple expression for T$_c$ of 2D ferromagnets by fitting classical Monte Carlo results for different lattice types. The expression they derived is a solely a function of the first principles obtained MAE and magnetic exchange constants. By calculating the magnetic exchange constants with DMC and the MAE from DFT+U, we were able to obtain an estimate of T$_c$ for the monolayers.

\subsection{\label{sec:results-jarvis}JARVIS Workflow}

\begin{figure*}
\begin{center}
\includegraphics[width=15cm]{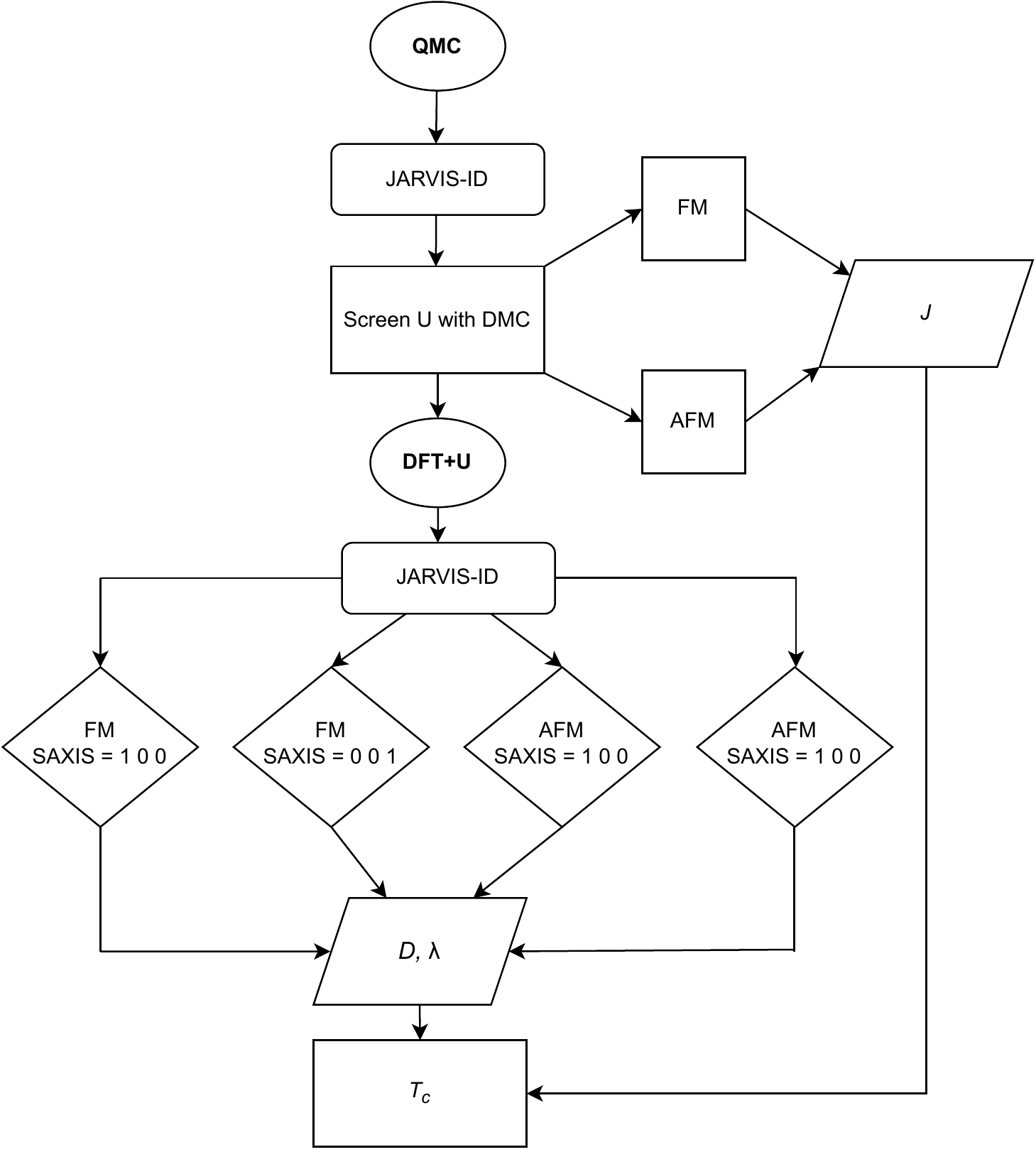}
\caption{The full high throughput workflow proposed in this work to obtain the accurate magnetic properties of a 2D system using a combination of DFT+U and QMC.  }
\label{flowchart}
\end{center}
\end{figure*}

The full form of the 2D model spin Hamiltonian \cite{Torelli_2018,Lado_2017} takes the form: 

\begin{equation}
{\cal {H}}= - \left( \sum_{i} D(S_i^z)^2 +\frac{J}{2}\sum_{i,i'}\vec{S}_i\cdot \vec{S}_{i'} + \frac{\lambda}{2}\sum_{i,i'}S_i^z S_{i'}^z \right)
\label{hamiltonian}
\end{equation}
where the sum over $i$ runs over the lattice of Cr atoms and $i'$ runs over the nearest Cr site of atom $i$ due to a strong magnetic moment localized on Cr atoms. Due to the fact that long-range interactions have previously been shown to die out in 2D CrI$_3$ \cite{Lado_2017}, we focused solely on the nearest neighbor interactions (our calculations consisted of a unit cell of two Cr atoms and six X atoms). From this unit cell (which is depicted in Fig. \ref{structure}), we adopt two magnetic orientations: ferromagnetic (FM), consisting of two spin up Cr atoms, and antiferromagnetic (AFM), consisting of one spin up Cr atom and one spin down Cr atom. This nearest neighbor AFM magnetic configuration is commonly referred to as the Neel configuration \cite{Lado_2017}. 

The first term in the spin Hamiltonian refers to the easy axis single ion anisotropy ($z$ chosen as the off-plane direction). The second term describes the Heisenberg isotropic exchange and the last term describes the anisotropic exchange. The sign convention is as follows: $J>0$ favors FM interactions, $D>0$ favors off-plane easy axis and $\lambda=0$ implies full isotropic exchange. We treat Eq. \ref{hamiltonian} classically, describing the spins $\vec{S}$ collinearly as either $S=S^x$ or $S=S^z$. By doing this, it possible to write the energy of the 4 possible magnetic ground states: i) ferromagnetic off-plane (FM,$z$), ii) antiferromagnetic off-plane (AFM,$z$), iii) ferromagnetic in-plane (FM,$x$), and iv) antiferromagnetic in-plane (AFM,$x$). The corresponding energy equations take the form:
\begin{equation}
\textrm{E}_{\textrm{FM},z} = -2S^2D-3S^2(J+\lambda)
\label{fmz}
\end{equation}

\begin{equation}
\textrm{E}_{\textrm{AFM},z} = -2S^2D+3S^2(J+\lambda)
\label{afmz}
\end{equation}

\begin{equation}
\textrm{E}_{\textrm{FM},x} = -3S^2J
\label{fmx}
\end{equation}

\begin{equation}
\textrm{E}_{\textrm{AFM},x} = +3S^2J
\label{afmx}
\end{equation}
where $S=3/2$. This model has been extensively used to model 2D magnetism \cite{Lado_2017,olsen-data}. However, the model assumes diagonal exchange interaction and does not include Dzyaloshinskii-Moriya interactions, Kitaev interactions, or higher order spin interactions. Recently, there have been attempts to modify this model with various types of spin interactions, but presently the role of the different terms are not completely understood. Most importantly, there are strong disagreements between calculations \cite{PhysRevB.101.060404,kitaev} and experiments \cite{PhysRevLett.124.017201} in regards to the magnitude of Kitaev interactions.

To obtain initial reference values for $J$, $\lambda$ and $D$, we performed self-consistent noncollinear DFT calculations in VASP (PAW). In these spin-orbit calculations, we rotated the easy axis by 90$^{\circ}$ and calculated the energy difference between the rotated and non-rotated configurations separately for FM and AFM. Since the magnetic anisotropy calculations are more difficult to converge than the total energy, we adjusted certain parameters to allow for careful convergence. Specifically, the cutoff energy was increased by 100 - 200 eV, the electronic convergence threshold was decreased to 1.0 $\times10^{-7}$ eV, symmetry was turned off, and the well-converged k-point grid obtained from JARVIS was used \cite{choudhary2019convergence}. Upon comparison to the magnetic anisotropy energy results from other recent studies \cite{https://doi.org/10.48550/arxiv.2205.00300,olsen-data,Torelli_2018} that used even higher convergence criteria (cutoff energy, k-points, supercell size), our results are in excellent agreement, which further demonstrates that our results are carefully converged. These four DFT calculations were automated using the JARVIS workflow, where four distinct total energy values were calculated for each material. This workflow was benchmarked for 2D CrI$_3$ (JVASP-76195), CrBr$_3$ (JVASP-6088), CrCl$_3$ (JVASP-76498) and CrF$_3$ (JVASP-153105) for using multiple flavors of DFT (discussed in detail in DFT Benchmarking section).

Although performing these four noncollinear DFT+U calculations is a robust method for determining the magnetic exchange and anisotropy parameters of a 2D system, these results can be systematically improved with QMC simulations. QMC can improve the magnetic property predictions in two ways. Firstly, one can variationally determine the optimal U value using DMC (discussed in further detail in the following section) and secondly, a statistical bound can be calculated for the $J$ parameter by performing DMC simulations for the FM and AFM phases separately. It is important to note that these QMC energies are collinear (spin-polarized), in contrast to the previous noncollinear (spin-orbit) DFT calculations. Currently, spin-orbit implementation is limited in DMC. For this reason, we neglect the $\lambda$ contribution when calculating $J$ with QMC using Eq. \ref{fmz} and \ref{afmz}, since such a contribution can only be obtained from spin-orbit calculations. This does not have a significant impact the final result for $J$, since $J >> \lambda$. As a result, we design a high-throughput workflow that allows us to variationally determine the optimal value of U using DMC, calculate a statistical bound on $J$ by performing collinear DMC calculations for the FM and AFM phases, and use that optimal determined U to perform DFT+U simulations and extract the anisotropy parameters ($D$, $\lambda$), with the end goal of using these parameters to accurately estimate the 2D critical temperature. A full schematic of this workflow is depicted in Fig. \ref{flowchart}.

It is possible to estimate the critical temperature using the analytical method outlined in Torelli and Olsen \cite{Torelli_2018} with our obtained values of $J$, $D$ and $\lambda$. In Torelli's work, classical Monte Carlo and Random Phase Approximation (RPA) simulations were used to derive a simple expression for T$_c$ that depends soley on lattice type and the ab-initio exchange coupling constants. The analytical function for $\textrm{T}_c$ takes the form:
\begin{equation}
T_c=T_c^{\textrm{Ising}}f(x)
\label{tc}
\end{equation}
with 
\begin{equation}
f(x)=\textrm{tanh}^{1/4}\left[\frac{6}{N_{nn}}\text{log}(1+\gamma x)\right]
\label{fx}
\end{equation}
where $N_{nn}$ is the number of nearest neighbors and $\gamma=0.033$ (dimensionless constant). $\textrm{T}_c^{\textrm{Ising}}$ is the critical temperature for the standard Ising model, which can be written as $\textrm{T}_c^{\textrm{Ising}}=S^2 J \tilde{T}_c/k_B$, where $\tilde{T}_c$ is the fitted dimensionless critical temperature (1.52 for honeycomb lattice). In cases where single ion anisotropy and anisotropic exchange are present, $x=\Delta/J(2S-1)$, where $\Delta$ is the spin gap:
\begin{equation}
\Delta=D(2S-1)+\lambda S N_{nn}.
\label{delta}
\end{equation}
Post-processing of the energies to determine $J$, $D$ and $\lambda$, and finally T$_c$ were also carried out within the JARVIS workflow.

\section{\label{sec:results}Results and Discussion}

\subsection{\label{sec:results-dft}DFT Benchmarking}

Prior to incorporating QMC calculations in the workflow (as depicted in Fig. \ref{flowchart}), we performed reference DFT calculations using VASP (PAW) to benchmark the magnetic properties of monolayer CrX$_3$ (X = I, Br, Cl, F). Performing these reference calculations in VASP is advantageous because it allows us to perform spin-orbit calculations and the DFT calculations come at a much lower cost than those performed with the NC pseudopotentials in QE (due to higher cutoff energy). Due to these advantages, we performed these calculations with a variety of local and semi-local density functionals such as PBE, LDA, SCAN, r2SCAN (with and without a U correction of 2 eV) and HSE06, including spin-orbit effects, to identify if such a choice made a real difference. Table \ref{tab:vasp1} depicts the values of $J$, $D$, $\lambda$ and T$_c$ calculated with each functional, using static geometry (depicted in Fig. \ref{structure}) and structures taken from the JARVIS-DFT database, where the geometry was optimized with vdW-DF-OptB88. A scatter plot of the data presented in Table \ref{tab:vasp1} is depicted in Fig. S4. The reason we did not perform this workflow with vdW-DF-OptB88 is because spin-orbit coupling is not compatible with vdW functionals such as vdW-DF-OptB88 \cite{Klime__2009}. Similar to previous DFT results, CrI$_3$ has the highest degree of MA when compared to the other 2D CrX$_3$ materials. This is due to the larger contribution of spin-orbit coupling that can be attributed to the I atoms in the cell. Depending on functional, a wide variety of $J$ is predicted for CrI$_3$, ranging from 1.98 meV to 4.22 meV. This drastic difference in magnetic exchange demonstrates the shortcomings of local/semi-local density functionals being used to deal with correlated electronic systems. This wide, functional dependent spread in $J$ is also observed for 2D CrBr$_3$, CrCl$_3$ and CrF$_3$ (see Table \ref{tab:vasp1}). Since the magnetic exchange is the driving force behind the magnitude of the Curie temperature, the estimation of $J$ is crucial for an accurate T$_c$. For example, the LDA $J$ value of 1.98 yields a T$_c$ of 27.09 K and the HSE06 $J$ value yields a T$_c$ of 48.63 K for CrI$_3$. With respect to the experimental value of T$_c$ = 45 K for CrI$_3$ \cite{cri3}, PBE+U, SCAN+U and HSE06 are closest.

In terms of the MA of CrI$_3$ ($D$ and $\lambda$), there is a relatively consistent trend between density functionals, with the exception of SCAN and SCAN+U. For SCAN and SCAN+U, the in-plane easy axis is favored ($D<0$) and for pure SCAN, a smaller value of $\lambda$ combined with this negative $D$ results in a negative spin gap (see Eq. \ref{delta}) and a nonphysical critical temperature in the Torelli and Olsen model. This negative spin gap behavior as a result of SCAN is not unique to CrI$_3$. As seen in Table \ref{tab:vasp1}, this occurs for CrBr$_3$ (calculated with SCAN+U) and CrF$_3$ (calculated with SCAN). This behavior can be due in part to the numerical instabilities of the SCAN functional when being used to calculate properties of complex and correlated systems. This is even more evident when comparing SCAN and SCAN+U findings to r2SCAN and r2SCAN+U (r2SCAN is meant to correct the numerical instabilities in SCAN \cite{r2scan}) results, where we only obtain positive spin gaps and finite T$_c$ (see Table \ref{tab:vasp1}). In recent literature, concerns about the performance of the SCAN functional for magnetic materials have been brought up, including the over-magnetization of transition metal solids, deeming SCAN inappropriate for open shell metallic ferromagnetic metals \cite{PhysRevB.100.041113,PhysRevB.100.045126}. However, SCAN has been reported to yield accurate properties across all bulk MnO$_2$ polymorphs \cite{PhysRevB.93.045132}. Regardless of these successes and concerns, we believe it is important to report these SCAN results as a benchmark to compare DMC and other density functionals to.

\begin{table}[]
\caption{\label{tab:vasp1}
Benchmarking noncollinear DFT and DFT+U (U = 2 eV) data calculated with various functionals (PBE, LDA, SCAN, r2SCAN, HSE06) and the VASP code with PAW pseudopotentials for CrX$_3$ monolayers. Values for $J$, $D$, $\lambda$ and T$_c$ are given for each functional and material. It is important to note that the geometry in these calculations is fixed to the geometry obtained from the JARVIS-2D DFT database (relaxed with vdW-DF-OptB88).}
\resizebox{9cm}{9cm}{%
\begin{tabular}{l|l|l|l|l}
\hline
\hline
CrI$_3$       &         &      &             &                        \\
\hline
Functional & $J$ (meV) & $D$ (meV) &$\lambda$ (meV) & T$_c$ (K)                    \\
\hline
PBE        & 2.83                        & 0.192                       & 0.173                            & 38.33                                         \\
\hline
PBE+U      & 3.70                        & 0.075                       & 0.161                            & 43.60                                         \\
\hline
LDA        & 1.98                        & 0.139                       & 0.126                            & 27.09                                         \\
\hline
LDA+U      & 2.80                        & 0.054                       & 0.147                            & 34.21                                         \\
\hline
SCAN       & 3.24                        & -0.129                      & 0.019                            & - \\
\hline
SCAN+U     & 4.06                        & -0.069                      & 0.216                            & 46.23                                         \\
\hline
r2SCAN     & 2.57                        & 0.145                       & 0.049                            & 29.00                                         \\
\hline
r2SCAN+U   & 3.08                        & 0.146                       & 0.044                            & 32.91                                         \\
\hline
HSE06      & 4.22                        & 0.068                       & 0.173                            & 48.63                                         \\
\hline
\hline
CrBr$_3$       &         &      &             &                        \\
\hline
Functional & $J$ (meV) & $D$ (meV) &$\lambda$ (meV) & T$_c$ (K)                    \\
\hline
PBE        & 2.14                        & 0.035                       & 0.032                            & 20.36                                         \\
\hline
PBE+U      & 2.85                        & 0.020                       & 0.032                            & 24.28                                         \\
\hline
LDA        & 1.44                        & 0.041                       & 0.031                            & 15.23                                         \\
\hline
LDA+U      & 2.45                        & 0.025                       & 0.031                            & 21.82                                         \\
\hline
SCAN       & 1.63                        & 0.083                       & -0.016                           & 13.40                                         \\
\hline
SCAN+U     & 1.83                        & 0.016                       & -0.052                           & - \\
\hline
r2SCAN     & 1.85                        & 0.037                       & 0.013                            & 16.21                                         \\
\hline
r2SCAN+U   & 1.98                        & 0.024                       & 0.016                            & 16.68                                         \\
\hline
HSE06      & 2.16                        & 0.011                       & 0.014                            & 16.17                                         \\
\hline
\hline
CrCl$_3$       &         &      &             &                        \\
\hline
Functional & $J$ (meV) & $D$ (meV) &$\lambda$ (meV) & T$_c$ (K)                    \\
\hline
PBE        & 1.20                        & 0.002                       & 0.007                            & 8.44                                          \\
\hline
PBE+U      & 2.04                        & -0.003                      & 0.009                            & 12.44                                         \\
\hline
LDA        & 0.49                        & 0.011                       & 0.002                            & 4.14                                          \\
\hline
LDA+U      & 1.87                        & 0.009                       & 0.002                            & 10.69                                         \\
\hline
SCAN       & 0.94                        & 0.032                       & -0.003                           & 7.64                                          \\
\hline
SCAN+U     & 1.33                        & 0.005                       & 0.004                            & 8.57                                          \\
\hline
r2SCAN     & 1.42                        & 0.007                       & 0.004                            & 9.33                                          \\
\hline
r2SCAN+U   & 1.80                        & 0.007                       & 0.004                            & 11.26                                         \\
\hline
HSE06      & 1.89                        & 0.006                       & 0.000                            & 8.91                                          \\
\hline
\hline
CrF$_3$       &         &      &             &                        \\
\hline
Functional & $J$ (meV) & $D$ (meV) &$\lambda$ (meV) & T$_c$ (K)                    \\
\hline
PBE        & 1.53                        & 0.055                       & 0.000                            & 13.45                                         \\
\hline
PBE+U      & 1.33                        & 0.053                       & -0.001                           & 11.91                                         \\
\hline
LDA        & 1.87                        & 0.068                       & 0.003                            & 16.83                                         \\
\hline
LDA+U      & 1.63                        & 0.065                       & 0.002                            & 14.91                                         \\
\hline
SCAN       & 1.41                        & -0.028                      & -0.044                           & - \\
\hline
SCAN+U     & 1.29                        & 0.048                       & -0.013                           & 8.97                                          \\
\hline
r2SCAN     & 1.66                        & 0.023                       & 0.000                            & 11.38                                         \\
\hline
r2SCAN+U   & 1.44                        & 0.050                       & -0.002                           & 12.22                                         \\
\hline
HSE06      & 1.02                        & 0.053                       & -0.001                           & 9.79                                         \\
\hline
\end{tabular}
}
\end{table}

\begin{table}[]
\caption{\label{tab:vasp2}
Benchmarking noncollinear DFT and DFT+U (U = 2 eV) data calculated with various functionals (PBE, LDA, SCAN, r2SCAN, HSE06) and the VASP code with PAW pseudopotentials for CrX$_3$ monolayers. Values for $J$, $D$, $\lambda$ and T$_c$ are given for each functional and material. For these calculations, the geometry was relaxed using each respective functional, using collinear spin-polarized DFT for the FM orientation of each material. After this preliminary relaxation, the geometry was fixed and used in the DFT workflow to calculate T$_c$. }
\resizebox{9cm}{9cm}{%
\begin{tabular}{l|l|l|l|l}
\hline
\hline
CrI$_3$       &         &      &             &                        \\
\hline
Functional & $J$ (meV) & $D$ (meV) &$\lambda$ (meV) & T$_c$ (K)                    \\
\hline
PBE        & 2.80                          & 0.211                         & 0.176                            & 38.42                                         \\
\hline
PBE+U      & 3.84                          & 0.077                         & 0.182                            & 46.06                                         \\
\hline
LDA        & 1.13  & 0.149 & 0.117    & 17.61                 \\
\hline
LDA+U      & 2.54                          & 0.083                         & 0.148                            & 32.52                                         \\
\hline
SCAN       & 2.73                          & 0.041                         & 0.071                            & 28.61                 \\
\hline
SCAN+U     & 4.00                          & -0.089                        & 0.167                            & 41.69                                         \\
\hline
r2SCAN     & 2.58                          & 0.142                         & 0.057                            & 29.48                                         \\
\hline
r2SCAN+U   & 3.17                          & 0.345                         & -0.077                           & 30.72                                         \\
\hline
\hline
CrBr$_3$       &         &      &             &                        \\
\hline
Functional & $J$ (meV) & $D$ (meV) &$\lambda$ (meV) & T$_c$ (K)                    \\
\hline
PBE        & 2.48                          & 0.041                         & 0.036                            & 23.53                                         \\
\hline
PBE+U      & 3.12                          & 0.016                         & 0.044                            & 27.52                                         \\
\hline
LDA        & 0.13  & 0.040 & 0.028    & 2.51                  \\
\hline
LDA+U      & 2.01                          & 0.022                         & 0.028                            & 18.37                                         \\
\hline
SCAN       & 1.73                          & 0.050                         & 0.018                            & 16.64                                         \\
\hline
SCAN+U     & 1.89                          & 0.061                         & 0.011                            & 17.56                 \\
\hline
r2SCAN     & 1.95                          & 0.037                         & 0.020                            & 17.76                                         \\
\hline
r2SCAN+U   & 2.01                          & 0.023                         & 0.020                            & 17.29                                         \\
\hline
\hline
CrCl$_3$       &         &      &             &                        \\
\hline
Functional & $J$ (meV) & $D$ (meV) &$\lambda$ (meV) & T$_c$ (K)                    \\
\hline
PBE        & 1.78                          & -0.022                        & 0.026                            & 13.49                                         \\
\hline
PBE+U      & 2.41                          & -0.029                        & 0.028                            & 16.77                                         \\
\hline
LDA        & -1.46 & 0.009 & 0.002    & - \\
\hline
LDA+U      & 1.26                          & 0.006                         & 0.001                            & 7.48                                          \\
\hline
SCAN       & 1.04                          & 0.041                         & 0.012                            & 10.64                                         \\
\hline
SCAN+U     & 1.26                          & 0.032                         & -0.023                           & - \\
\hline
r2SCAN     & 1.57                          & -0.012                        & 0.024                            & 12.67                                         \\
\hline
r2SCAN+U   & 1.92                          & -0.018                        & 0.026                            & 14.63                                         \\
\hline
\hline
CrF$_3$       &         &      &             &                        \\
\hline
Functional & $J$ (meV) & $D$ (meV) &$\lambda$ (meV) & T$_c$ (K)                    \\
\hline
PBE        & 1.74                          & 0.058                         & 0.001                            & 15.12                                         \\
\hline
PBE+U      & 1.44                          & 0.057                         & -0.001                           & 12.86                                         \\
\hline
LDA        & 1.12                          & 0.067                         & 0.003                            & 11.41                                         \\
\hline
LDA+U      & 1.55                          & 0.061                         & -0.001                           & 13.86                                         \\
\hline
SCAN       & 1.49                          & 0.025                         & -0.007                           & 8.52                  \\
\hline
SCAN+U     & 1.22                          & 0.044                         & -0.006                           & 9.76                                          \\
\hline
r2SCAN     & 1.68                          & 0.037                         & 0.008                            & 14.38                                         \\
\hline
r2SCAN+U   & 1.42                          & 0.049                         & 0.020                            & 14.45                                         \\
\hline
\end{tabular}
}
\end{table}

In addition to performing this DFT/DFT+U workflow for a fixed geometry, we investigated the geometry dependence on the magnetic exchange and anisotropy. We did so by first relaxing the FM orientation of each structure with each respective functional (PBE, LDA, SCAN, r2SCAN, and +U for each) using spin-polarized DFT (as opposed to spin-orbit DFT, which comes at a much higher computational cost for geometric relaxation calculations). The relaxation using HSE06 was omitted due to the high computational expense of such calculations. Once the relaxed FM geometry was obtained, we fixed this geometry and performed the same noncollinear DFT (as Table \ref{tab:vasp1}) workflow to obtain the magnetic constants. The results of these calculations are presented in Table \ref{tab:vasp2} and the relaxed structural parameters are given in Table S2. A scatter plot of the data presented in Table \ref{tab:vasp2} is depicted in Fig. S5.

For the most part, geometry has little effect on the magnetic properties of each monolayer, which implies that the functional dependence is stronger. When comparing Table \ref{tab:vasp1} and Table \ref{tab:vasp2} for 2D CrI$_3$, we observe that the $J$ parameters and T$_c$ values are consistent between whether the geometry is relaxed or the geometry is fixed to the vdW-DF-OptB88 (JARVIS) structure. The only exception of this is the LDA relaxed geometry, where the lattice parameters are severely underestimated (see Table S2) for CrI$_3$ and subsequently the values for $J$ and T$_c$ are much lower than expected (1.13 meV and 17.16 K). This LDA trend also occurs for monolayer CrBr$_3$, CrCl$_3$ and CrF$_3$. For CrCl$_3$, the lattice compression from the LDA relaxation causes a magnetic phase transition from FM favorable to AFM favorable (see negative $J$ value in Table \ref{tab:vasp2}). This compression-induced phase transition has been reported theoretically for 2D CrCl$_3$ in Dupont et al. \cite{PhysRevLett.127.037204}. Changes in the MA energy with respect to geometry are most prevalent for 2D CrCl$_3$, where we see a shift from out-of- plane easy axis ($D>0$) to in-plane easy axis ($D<0$) favorability for PBE and r2SCAN. This sign change in $D$ occurs while relaxing with the respective functionals for CrCl$_3$, but there is only a small shift in T$_c$ (between Table \ref{tab:vasp1} and Table \ref{tab:vasp2}) due to the fact that $J$ is the driving force behind the critical temperature. For these reasons, we proceeded to use the vdW-DF-OptB88 relaxed geometry for subsequent QMC calculations in the next section. Further justification of using the vdW-DF-OptB88 geometry for QMC calculations stems from the fact that the geometry of 2D CrI$_3$ calculated with vdW-DF-OptB88 is in identical statistical agreement with the geometry calculated from a previous DMC study from Staros et al. \cite{staros} (see Table S2). In addition, the lattice constant obtained with vdW-DF-OptB88 for 2D CrCl$_3$ is in the closest agreement with the bulk lattice constant of layered CrCl$_3$\cite{doi:10.1063/1.1725428} (see also Table S2). 

Although the structural parameters impact the magnetic properties of CrCl$_3$ more significantly than CrI$_3$, CrBr$_3$ and CrF$_3$ (see difference between Table \ref{tab:vasp1} and Table \ref{tab:vasp2}), overall the functional dependence is stronger than the structural dependence. For comparison purposes, we present literature results \cite{Lado_2017,Torelli_2018,PhysRevLett.127.166402,Kumar_Gudelli_2019,kitaev,olsen-data,PhysRevB.98.144411,C5TC02840J,Pizzochero_2020} for all four monolayers calculated with a variety of computational methods in Table S3. Similar to our own results, we see a large variability in $J$ and T$_c$ with respect to the computational method used, but we observe the same overall trends between each material. This is especially prevalent in works such as the one by Pizzochero et al. \cite{Pizzochero_2020}, which benchmarks the properties of CrI$_3$ for a variety of DFT methods and multi-reference configuration interaction (MRCI) theory.

This high-throughput benchmarking with VASP is an important preliminary step in identifying the key areas where higher order correlated methods such as QMC can be used to improve electronic and magnetic property prediction. The next section will detail the process of incorporating QMC calculations into the workflow (as depicted in Fig. \ref{flowchart}) and provide a deeper analysis of our QMC results.

\subsection{\label{sec:results-qmc}QMC Results}

\begin{figure}
\begin{center}
\includegraphics[width=6cm]{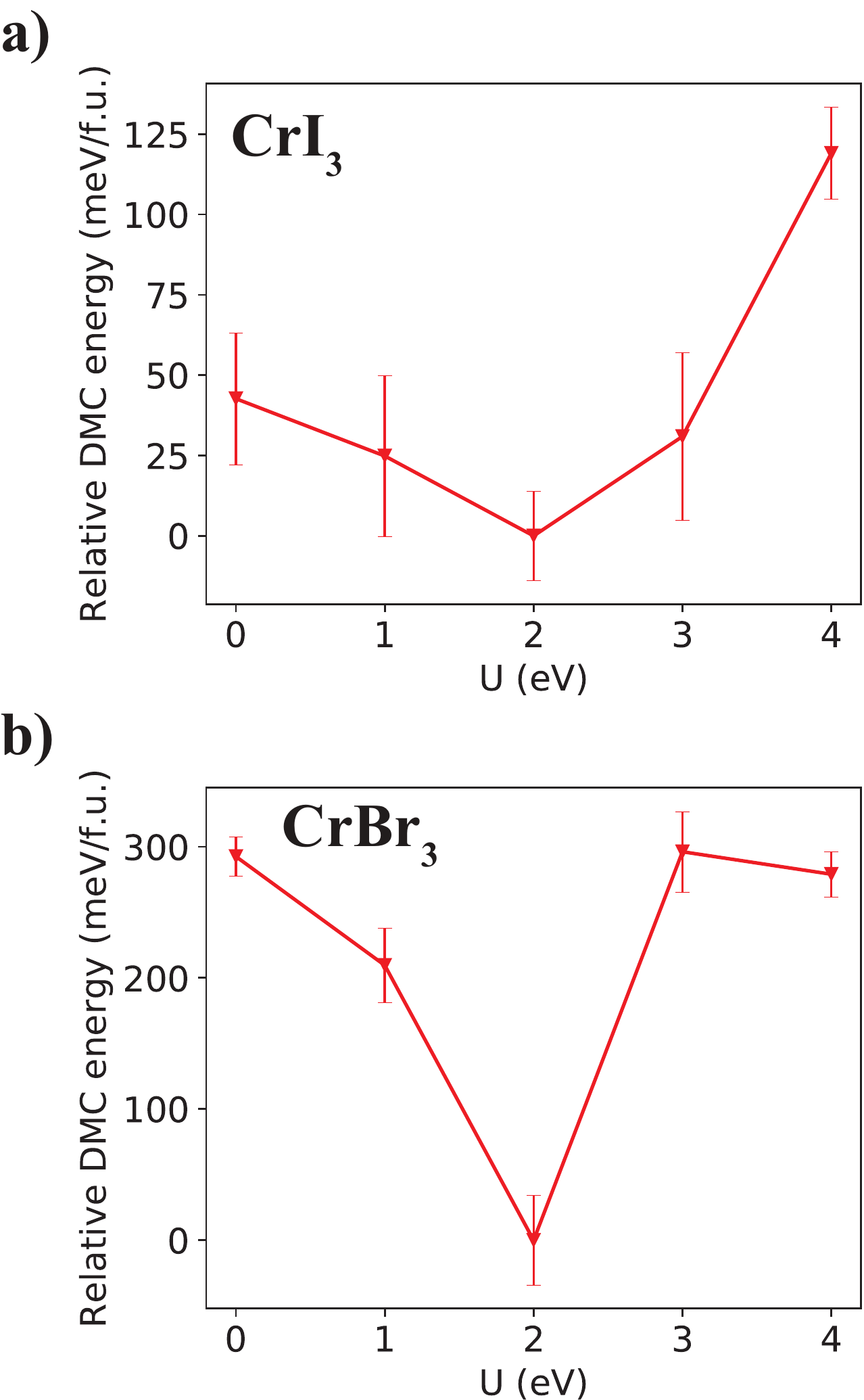}
\caption{DMC
calculated total energies and subsequent standard error about the mean (indicated by error bars) of a 16-atom supercell (normalized per formula unit (f.u.)) of the ferromagnetic orientation of 2D a) CrI$_3$ and b) CrBr$_3$ calculated as a function of the U parameter used to variationally determine the optimal trial wave function. For
convenience of presentation, the DMC energies are shifted by the lowest DMC energy obtained at the appropriate U value (U = 2 eV for CrI$_3$ and CrBr$_3$).   }
\label{u-tune}
\end{center}
\end{figure}

QMC calculations in this study (which consist of a VMC calculation to optimize the Jastrow parameters of the wavefunction, followed by a DMC calculation) require a well converged trial wavefunction that is created via a DFT calculation. As opposed to the VASP DFT (PAW) calculations in the previous section, the DFT calculations detailed in this section (for additional benchmarking and creating the trial wavefunction) are performed using NC pseudopotentials (for additional information see Computational Methods). DMC has the property of zero-variance, which means that as the trial wave function approaches the exact ground state (exact nodal surface), the statistical fluctuations in the energy reduce to zero \cite{RevModPhys.73.33}. There have been reported instances where various sophisticated and expensive methods have been used to optimize the nodal surface of the trial wave function \cite{PhysRevB.48.12037,PhysRevB.58.6800,PhysRevE.74.066701,PhysRevLett.104.193001}. Similar to other DMC studies of magnetic materials \cite{PhysRevX.4.031003,PhysRevMaterials.5.064006,PhysRevMaterials.3.124414,PhysRevMaterials.2.085801,staros,mno2-qmc}, we employed a PBE+U approach where the Hubbard U value was used as a variational parameter to optimize the nodal surface using DMC. Since we can determine the optimal U parameter variationally using DMC, it makes our final results more reliable than solely using DFT+U, where the U parameter is  arbitrarily chosen or fitted to experimental data. In addition, the DMC determined U value can be used for subsequent DFT+U calculations, giving a more reliable and less costly method to compute correlated material properties. 

Fig. \ref{u-tune} depicts the total energies of a 16-atom supercell (normalized per formula unit (f.u.)) of the FM orientation of 2D a) CrI$_3$ and b) CrBr$_3$ while Fig. S6 depicts a) CrCl$_3$, and b) CrF$_3$ calculated as a function of the U parameter, with the goal of variationally determining the optimal trial wave function. For
convenience of presentation, the DMC energies are shifted by the lowest DMC energy obtained at the appropriate U value (U = 2 eV for CrI$_3$ and CrBr$_3$, U = 1 eV for CrCl$_3$ and U = 4 eV for CrF$_3$). Unsurprisingly, the U value that yields the lowest energy for 2D CrI$_3$ is 2 eV, which is similar to the result obtained by Staros et al. \cite{staros}. U = 2 eV also yields the lowest energy for CrBr$_3$, but the difference in energy between U = 2 eV and other values of U is much larger than that of CrI$_3$, indicating that the U dependence on the trial wavefunction is stronger for 2D CrBr$_3$. As seen in Fig. S6 a) and b), U = 1 eV produces an optimal wavefunction for CrCl$_3$ and U = 4 eV produces an optimal wavefunction for CrF$_3$. Although these U values differ from the U = 2 eV value for CrI$_3$ and CrBr$_3$, the energies at the minimum points are statistically identical to the energy produced at U = 2 eV for CrCl$_3$ and CrF$_3$.

\begin{figure}
\begin{center}
\includegraphics[width=5cm]{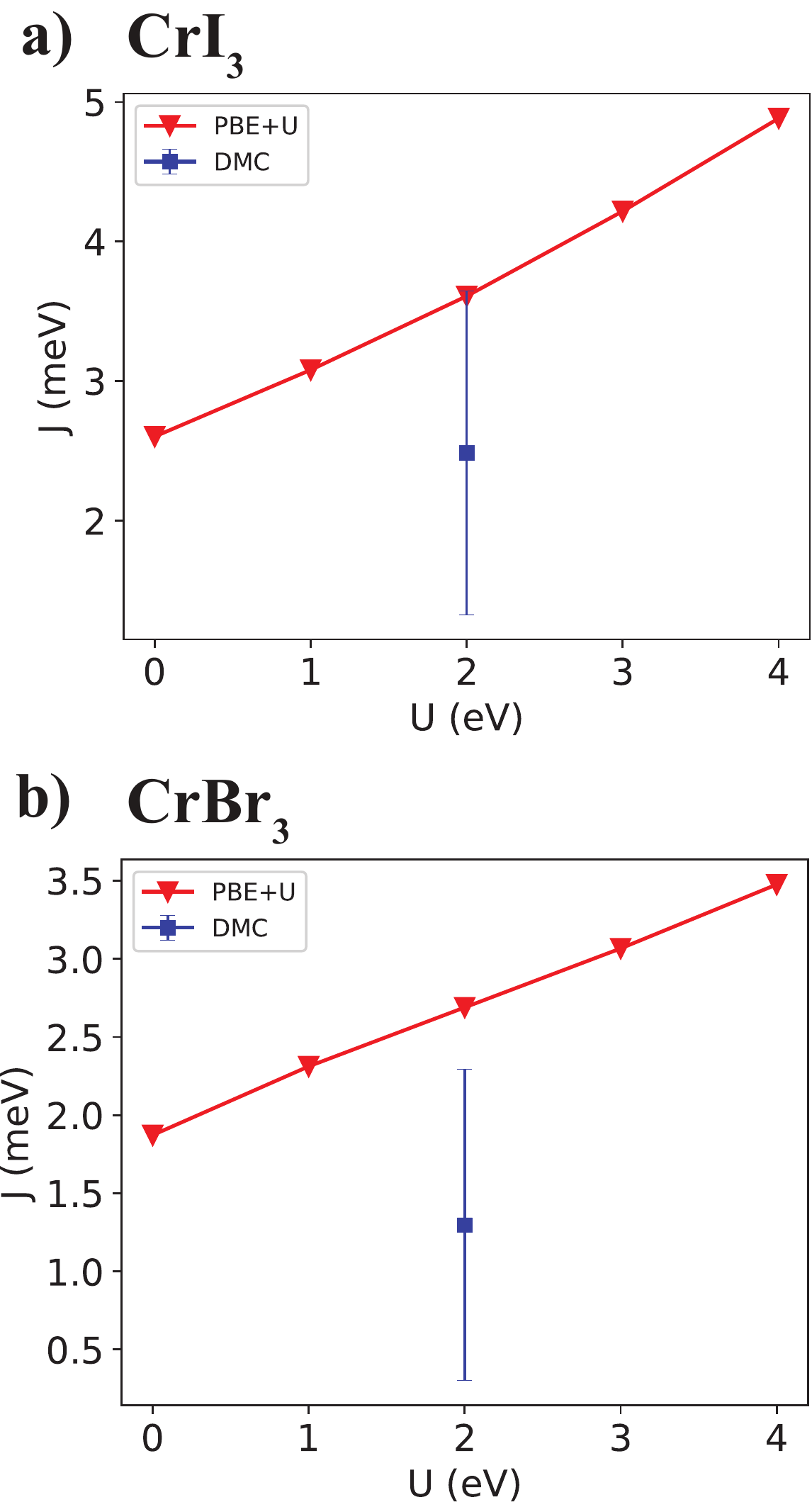}
\caption{The values of $J$ for 2D a) CrI$_3$ and b) CrBr$_3$ as a function of U calculated with PBE+U (red) and DMC (blue, using PBE+2 to create the trial wavefuction). Both methods used NC pseudopotentials in the calculation and a 48 atom supercell was used for each material. The standard error about the mean is indicated by error bars for DMC in blue.     }
\label{j-full}
\end{center}
\end{figure}

Due to the fact that 2D CrI$_3$ and CrBr$_3$ have the highest predicted T$_c$ out of the four CrX$_3$ monolayers (at the DFT level, see Table \ref{tab:vasp1} and Table \ref{tab:vasp2}), and the fact that the largest discrepancy in predicted $J$ and T$_c$ occurs for CrI$_3$ and CrBr$_3$, we decided to focus primarily on these two materials for the remainder of this study. Therefore, we ran monolayer CrI$_3$ and CrBr$_3$ through the full workflow depicted in Fig. \ref{flowchart}, with the goal of determining a statistical bound on $J$ and T$_c$. In order to compute $J$, we had to perform calculations for both monolayers in the FM and AFM (Neel) configurations. We performed these calculations using NC pseudpotentials (as previously mentioned) and based on our results presented in Fig. \ref{u-tune} a) and b), we created the trial wavefunction with PBE+U (U = 2 eV). For the sake of completeness, we also determined that U = 2 eV yields the optimal wavefunction for the AFM phase of CrI$_3$ and CrBr$_3$ (see Fig. S7), and this was used for all subsequent DMC calculations of the AFM phase. In addition, we performed these QMC simulations at a supercell size of 48 atoms (see Fig. S3 for convergence details). Fig. \ref{j-full} depicts the calculated $J$ parameter of 2D a) CrI$_3$ and b) CrBr$_3$ using PBE+U (red triangle) and DMC (blue square). It is important to note that the DMC simulations required to achieve an estimate for $J$ (both FM and AFM calculations) require significant computational resources. Specifically, the DMC estimate of $J$ required $\sim$ 1.0$\times10^6$ seconds/node to properly reduce the uncertainty (further details in supporting information). With more substantial computational resources beyond our capabilities, the uncertainty could be further reduced. As expected with PBE+U, as the U value increases from U = 0 eV to 4 eV, the $J$ parameter also increases linearly. In comparison to the PBE+U values, the average DMC calculated $J$ for CrI$_3$ falls slightly under that of PBE+2 value, with the error bar overlapping with the DFT values at U = [0, 1, 2] eV. For 2D CrBr$_3$, the average value of $J$ falls about 1 meV under the PBE+2 value, with the error bar overlaping with the DFT values at U = 0 and 1 eV. We report a DMC $J$ value of 2.49(1.16) meV for 2D CrI$_3$ and a value of 1.30(1.00) meV for 2D CrBr$_3$. This average DMC value of 2.49(1.16) meV for CrI$_3$ is in excellent agreement with the MRCI calculated value of 2.88 meV reported in Pizzochero et al. \cite{Pizzochero_2020} (see Table S3).

\begin{table}
\caption{\label{magneticdata} This table depicts the calculated $J$ with PBE+U and DMC (both using NC pseudopotentials) and the Ising temperature ($\textrm{T}_c^{\textrm{Ising}}$) and Curie temperatures ($\textrm{T}_c$) calculated with the Torelli and Olsen model. Columns 4 - 8 use various functionals for the MA from the VASP (PAW) calculations from the DFT Benchmarking section (at U = 2 eV) indicated by the superscript on $\textrm{T}_c$. The associated uncertainty of the DMC quantities are given in parenthesis.       }
\resizebox{\columnwidth}{!}{%
\begin{tabular}{l|l|l|l|l|l|l|l}
\hline
\hline
CrI$_3$        &        &        &              &              &                           &                   &               \\
\hline
\hline
Method      & $J$ (meV) & $\textrm{T}_c^{\textrm{Ising}}$ (K)  & $\textrm{T}_c^{\textrm{PBE+2}}$ (K) & $\textrm{T}_c^{\textrm{LDA+2}}$ (K) & $\textrm{T}_c^{\textrm{SCAN+2}}$  (K)       & $\textrm{T}_c^{\textrm{r2SCAN+2}}$ (K) & $\textrm{T}_c^{\textrm{HSE06}}$ (K) \\
\hline
PBE+0       & 2.60                        & 103.26                       & 33.46                               & 32.40                               & 33.10 & 29.00                                  & 33.81                               \\
\hline
PBE+1       & 3.08                        & 122.23                       & 37.98                               & 36.78                               & 37.57 & 32.91                                  & 38.38                               \\
\hline
PBE+2       & 3.61                        & 143.23                       & 42.79                               & 41.42                               & 42.32 & 37.07                                  & 43.23                               \\
\hline
PBE+3       & 4.22                        & 167.39                       & 48.10                               & 46.57                               & 47.58 & 41.67                                  & 48.60                               \\
\hline
PBE+4       & 4.88                        & 193.84                       & 53.70                               & 51.99                               & 53.12 & 46.52                                  & 54.26                               \\
\hline
DMC & 2.49(1.16)                        & 98.65(46.02)                        & 32.34(10.78)                               & 31.31(10.43)                               & 31.98(10.66) & 28.02(9.33)                                  & 32.67(10.89)                               \\
\hline
\hline
CrBr$_3$        &        &        &              &              &                           &                   &               \\
\hline
\hline
Method      & $J$ (meV) & $\textrm{T}_c^{\textrm{Ising}}$ (K)  & $\textrm{T}_c^{\textrm{PBE+2}}$ (K) & $\textrm{T}_c^{\textrm{LDA+2}}$ (K) & $\textrm{T}_c^{\textrm{SCAN+2}}$  (K)       & $\textrm{T}_c^{\textrm{r2SCAN+2}}$ (K) & $\textrm{T}_c^{\textrm{HSE06}}$ (K) \\
\hline
PBE+0       & 1.87                        & 74.31                        & 17.70                               & 17.84                               & -                         & 15.98                                  & 14.52                               \\
\hline
PBE+1       & 2.31                        & 91.76                        & 20.74                               & 20.90                               & -                         & 18.72                                  & 17.01                               \\
\hline
PBE+2       & 2.69                        & 106.74                       & 23.23                               & 23.41                               & -                         & 20.97                                  & 19.05                               \\
\hline
PBE+3       & 3.07                        & 121.72                       & 25.63                               & 25.84                               & -                         & 23.14                                  & 21.03                               \\
\hline
PBE+4       & 3.48                        & 137.97                       & 28.16                               & 28.38                               & -                         & 25.42                                  & 23.10                               \\
\hline
DMC  & 1.30(1.00)                        & 51.45(39.52)                        & 13.43(7.17)                               & 13.54(7.23)                               & -                         & 12.13(6.47)                                  & 11.02(5.88)        \\
\hline
\end{tabular}
}
\end{table}

In order to understand the implications of the DMC results presented in Fig. \ref{j-full}, we went a step further and calculated the Ising temperature (as described in the JARVIS Workflow section) and Curie temperature (T$_c$, calculated with the Torelli and Olsen model). These results are presented in Table \ref{magneticdata}. It is important to reiterate that the PBE+U and DMC $J$ (and subsequent $\textrm{T}_c^{\textrm{Ising}}$) values are determined from spin-polarized calculations and NC pseudopotentials, while the MA that is used in the Torelli and Olsen model to determine T$_c$ is carried out using VASP with PAW pseudopotentials (from the DFT Benchmarking section). This combination of methods (outlined in Fig. \ref{flowchart}) allows us to put a statistical bound on T$_c$ for a given magnetic material. As seen in Table \ref{magneticdata}, we present T$_c$ calculated with various values of $J$ (rows: PBE+U with varying U and DMC at U = 2 eV) and various values for the MA parameters (columns: using various functionals at the optimal U of 2 eV, indicated by the superscript on T$_c$). It is clear from Table \ref{magneticdata} that $J$ is more of a driving force behind the variability of T$_c$ than the MA. The largest difference in anisotropy (and therefore T$_c$) occurs for r2SCAN+2 for CrI$_3$ and r2SCAN+2 and HSE06 for CrBr$_3$, with respect to other functionals used to calculate anisotropy. Similarly to the data reported in Table \ref{tab:vasp1}, we observe a nonphysical T$_c$ for CrBr$_3$ with MA calculated with SCAN+2 (due to a negative spin gap). For 2D CrI$_3$, we obtain a statistical bound of T$_c$ = 43.56 K using the $J$ obtained from DMC and the anisotropy parameters obtained from HSE06, which is in excellent agreement with the measured value of 45 K \cite{cri3}. For 2D CrBr$_3$, we obtain a maximum value of T$_c$ = 20.78 K using the $J$ obtained from DMC and the anisotropy parameters obtained from LDA+2. For the sake of completeness, we calculated the T$_c$ with a fixed $J$ obtained from DMC and the anisotropy ($D$ and $\lambda$) from a wider range DFT functionals, resulting in a sweep of possible values for T$_c$ (a more detailed extension of Table \ref{magneticdata}). This thorough sensitivity analysis of the DMC-DFT calculated critical temperature is given in Table S4, where we observe that the different values of anisotropy have a $\sim$ 20 $\%$ impact on the variability of the critical temperature for CrI$_3$ and CrBr$_3$. Interestingly, anisotropy obtained from PBE (no U correction) results in a slightly higher critical temperature for both monolayers (and bare LDA for CrBr3), but we find our maximum value estimates of T$_c$ = 43.56 K and T$_c$ = 20.78 K (for CrI$_3$ and CrBr$_3$ respectively) to be more reliable since the anisotropy was obtained from hybrid and Hubbard corrected functionals. Although these critical temperatures are far below room temperature, it has been demonstrated that Tc can be increased by applying strain \cite{fm-mno2} or by changing the monolayer substrate \cite{vse2}.

\begin{figure*}
\begin{center}
\includegraphics[width=15cm]{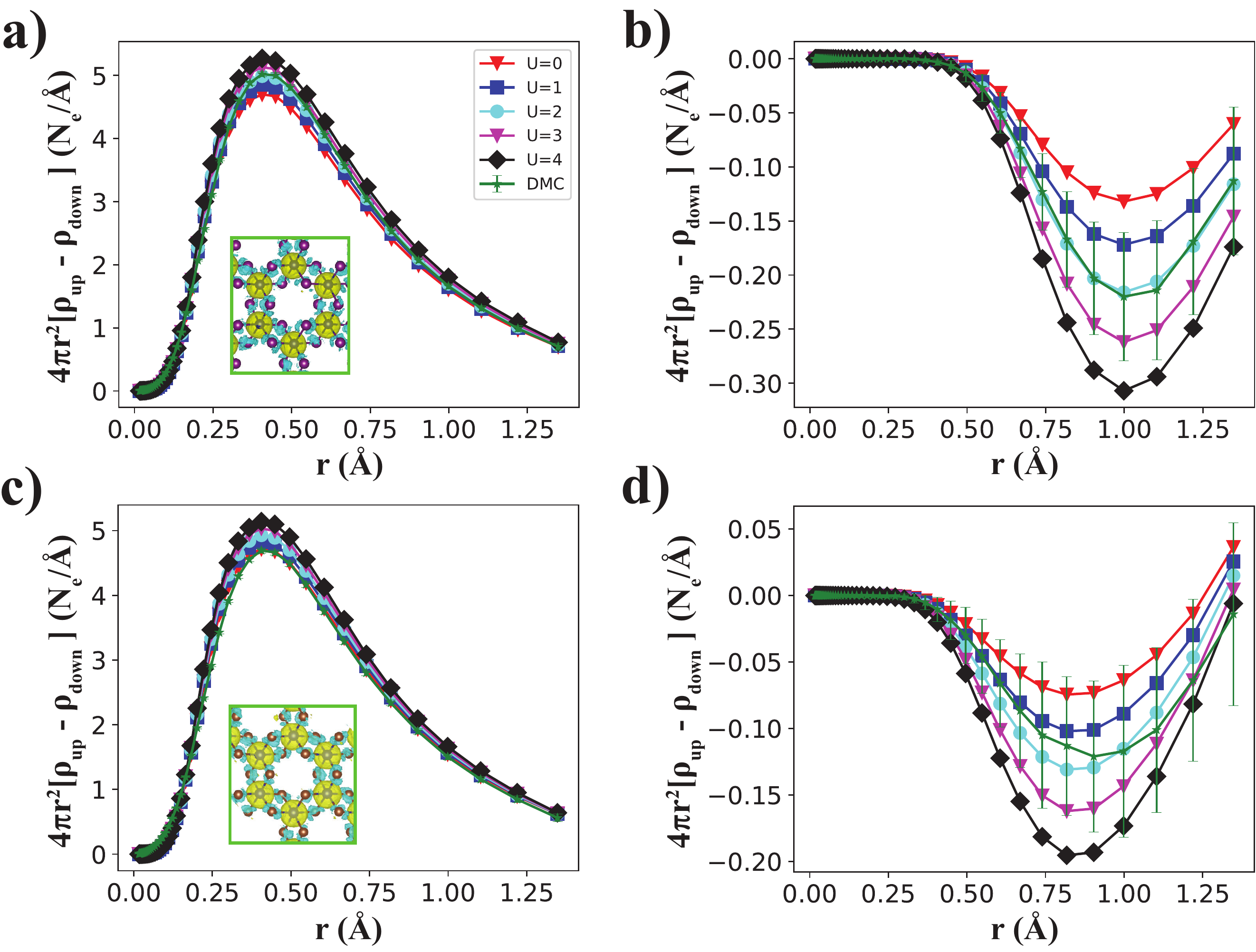}
\caption{The radially averaged spin density ($\rho_{up}$ - $\rho_{down}$) as a function of distance, calculated with DMC and PBE+U (U = [0, 1, 2, 3, 4] eV) of a) Cr and b) I for 2D CrI$_3$ and c) Cr and d) Br for 2D CrBr$_3$. The inset of a) and c) depicts the spin isosurface density of CrI$_3$ and CrBr$_3$ respectively, where the isosurface value was set to 5 x 10$^{-5}$ e/\AA$^{3}$. The standard error about the mean for DMC is indicated by error bars in green. }
\label{spindens}
\end{center}
\end{figure*}

\begin{table}[]
\caption{\label{magmom} Site-averaged atomic magnetic moments ( in $\mu_B$) of Cr and I for 2D CrI$_3$ and Cr and Br for 2D CrBr$_3$, estimated by integrating the spin density for DMC and PBE+\textrm{U} results. DMC uncertainties are given in the parenthesis.}
\begin{tabular}{l|l|l}
\hline
\hline
 CrI$_3$      &     & \multicolumn{1}{l}{}         \\
 \hline
 \hline
Method & M$_{\textrm{Cr}}$ ($\mu_B$)      &  M$_{\textrm{I}}$ ($\mu_B$)        \\
\hline
PBE+0  & 3.06                        & -0.08                        \\
\hline
PBE+1  & 3.15                        & -0.1                         \\
\hline
PBE+2  & 3.25                        & -0.13                        \\
\hline
PBE+3  & 3.34                        & -0.16                        \\
\hline
PBE+4  & 3.43                        & -0.18                        \\
\hline
DMC    & 3.21(5) & -0.13(5) \\
\hline
\hline
 CrBr$_3$      &   &        \\
 \hline 
 \hline
  Method     &  M$_{\textrm{Cr}}$ ($\mu_B$)      & M$_{\textrm{Br}}$ ($\mu_B$)         \\
       \hline
PBE+0  & 3.01                        & -0.03                        \\
\hline
PBE+1  & 3.07                        & -0.05                        \\
\hline
PBE+2  & 3.14                        & -0.07                        \\
\hline
PBE+3  & 3.2                         & -0.09                        \\
\hline
PBE+4  & 3.26                        & -0.11                        \\
\hline
DMC    & 2.96(5) & -0.07(5) \\
\hline
\end{tabular}
\end{table}

As an additional benchmark, we extracted the total charge density from our DMC simulations of 2D CrI$_3$ and CrBr$_3$ (using a trial wavefunction at U = 2 eV and 48 atom supercell). From the total charge densities, we were able to determine the spin densities ($\rho_{\textrm{up}}$ - $\rho_{\textrm{down}}$), which are depicted in the insets of Fig. a) and c) for CrI$_3$ and CrBr$_3$ respectively. From this many-electron DMC approach, we observe that for both materials, the Cr atoms are highly spin-polarized while the I and Br atoms are slightly polarized antiparallel with respect to the Cr atoms. We went a step further by plotting the radial averaged densities as a function of distance for Cr and I separately for CrI$_3$ and Cr and Br separately for CrBr$_3$. This gives us the spatial variations in total charge density (Fig. S8) and spin density (Fig. \ref{spindens}). We benchmarked these radially averaged densities with PBE+U (U = [0, 1, 2, 3, 4] eV) using NC pseudopotentials in QE. 

Fig. S8 a) and c) depicts the radially averaged total charge density of the Cr atoms as a function of distance for CrI$_3$ and CrBr$_3$ respectively. We observe that while the PBE+U results are almost identical for the Cr atoms for both materials, the total charge density of Cr is overestimated (mostly around the peak) with respect to the DMC total density. Although this PBE+U overestimation occurs for both CrI$_3$ and CrBr$_3$, it is much more apparent for CrBr$_3$ (see Fig. S8 c)). For I and Br (Fig. S8 b) and d)), the difference between the PBE+U and DMC total charge density is negligible. The larger discrepancy between DMC and PBE+U for the Cr atom (for both CrI$_3$ and CrBr$_3$) near the radial density peak (peak of d orbital) is due in part to the fact that DFT functionals tend to unsuccessfully capture 3$d$ orbitals. Although this sizeable difference between PBE+U and DMC occurs for the total charge densities, it has been reported that various DFT methods can give a more accurate description of the spin density than the total charge density \cite{PhysRevMaterials.1.065408}. Fig. \ref{spindens} depicts the radially averaged spin densities for each atom of 2D CrI$_3$ and CrBr$_3$. For CrI$_3$ (Fig. \ref{spindens} a) and b)), we observe excellent agreement between the DMC and PBE+2 spin densities for Cr and I, indicating that the PBE+2 method does not only reproduce a correct T$_c$ with respect to DMC, but also correct spin density. In contrast, the results for CrBr$_3$ are slightly different. Although it was determined that PBE+2 yields the most optimal wavefunction for DMC (Fig. \ref{u-tune} b)), the DMC spin density of Cr in CrBr$_3$ is closest to the PBE+0 result (see Fig. \ref{spindens} c)). This is consistent with the trend of the DMC calculated $J$ (and T$_c$) overlapping more closely with PBE+0 for CrBr$_3$ (see Fig. \ref{j-full} b) and Table \ref{magneticdata}). As for the Br atom in CrBr$_3$, the DMC spin density is within the margin of error of that calculated with PBE+U for U = [0, 1, 2 and 3] eV. Most importantly for both materials (Fig. \ref{spindens} b) and d)), the antiparallel polarization of I and Br with respect to Cr is present for DMC and all PBE+U results. 

We went one step further and estimated the site-averaged atomic magnetic moments per Cr and I for 2D CrI$_3$ and Cr and Br for CrBr$_3$ by integrating the spin densities depicted in Fig. \ref{spindens}. These tabulated magnetic moments are presented in Table \ref{magmom}. The results of Table \ref{magmom} are consistent with the spin density results presented in Fig. \ref{spindens}, where we see that the DMC calculated magnetic moment for CrI$_3$ is closest to PBE+2 and the DMC calculated magnetic moment of CrBr$_3$ is closest to PBE+0. Since PBE+0 produces results closest to DMC, we decided to recalculate the T$_c$ using the anisotropy from PBE+U (U = 0 eV). We find that this increases the DMC maximum value to T$_c$ = 21.39 K, which is about 1 K larger than previously reported (see Table S4). By analyzing and integrating the spin densities, we obtain a clear picture of how the magnetization of each ion depends on the computational method used. These results serve as a many-body theoretical benchmark for the magnetic properties of 2D CrI$_3$ and CrBr$_3$ and give information on how to assess the accuracy of DFT calculations with various Hubbard corrections.

\section{\label{sec:conc}Conclusion}

In this work we designed and applied a workflow that combines DFT+U, QMC (VMC, DMC) and analytical models to estimate a statistical bound for the critical temperature of a 2D magnetic system. Such a workflow is intended to be integrated into the JARVIS framework. We chose monolayer CrX$_3$ (X = I, Br, Cl, F) as a case study since they have been experimentally realized and have a finite critical temperature. After extensive DFT+U benchmarking with several functionals, we deemed that 2D CrI$_3$ and CrBr$_3$ were worthwhile to run through the more computationally expensive DFT+U and QMC workflow, due to their higher T$_c$ and higher degree of disagreement between DFT functionals. After variationally determining the optimal wavefunction for DMC (Hubbard U value used in the DFT wavefunction generation), we calculated maximum value of 43.56 K for the T$_c$ of CrI$_3$ and of 20.78 K for the T$_c$ of CrBr$_3$. We also extracted the spin-density from our DMC results for Cr and I atoms separately for CrI$_3$ and Cr and Br atoms separately for CrBr$_3$ and provide a detailed comparison with DFT+U. In terms of the workflow, this procedure can be used for the investigation of future 2D magnetic systems that have a higher degree of complexity and electron correlation, such as transition metal oxides. The findings of this specific case study show the successes of the DMC method when applied to a 2D magnetic system and provide a many-body theoretical benchmark for CrX$_3$ monolayers that will guide experimentalists in characterizing 2D magnets.

\section{Data Availability Statement}
The data from the present work is available at https://github.com/wines1/CrX3-QMC. 

 \section{Code Availability Statement}
Software packages mentioned in the article can be found at https://github.com/usnistgov/jarvis.

 \section{Competing interests}
The authors declare no competing interests.

  \section{Supporting Information}
 Energy cutoff convergence, k-point convergence, DMC timestep convergence, finite-size convergence, tabulated geometric properties, scatter plot of DFT computed magnetic exchange and anisotropy, tabulated comparison to literature of magnetic properties, supplemental figures for nodal surface optimization, table that illustrates sweep of T$_c$ values calculated from DMC and DFT, total charge density.
 
\section{Acknowledgments}
The authors thank the National Institute of Standards and Technology for funding, computational, and data-management resources. K.C. thanks the computational support from XSEDE (Extreme Science and Engineering Discovery Environment) computational resources under allocation number TG-DMR 190095. Contributions from K.C. were supported by the financial assistance award 70NANB19H117 from the U.S. Department of Commerce, National Institute of Standards and Technology. The authors thank Dr. Can Ataca and Dr. Kayahan Saritas for fruitful discussions.

\section*{References}
\bibliographystyle{plain}
\bibliography{main}% Produces the bibliography via BibTeX.

\end{document}

\maketitle

\begin{figure}
\begin{center}
\includegraphics[width=14cm]{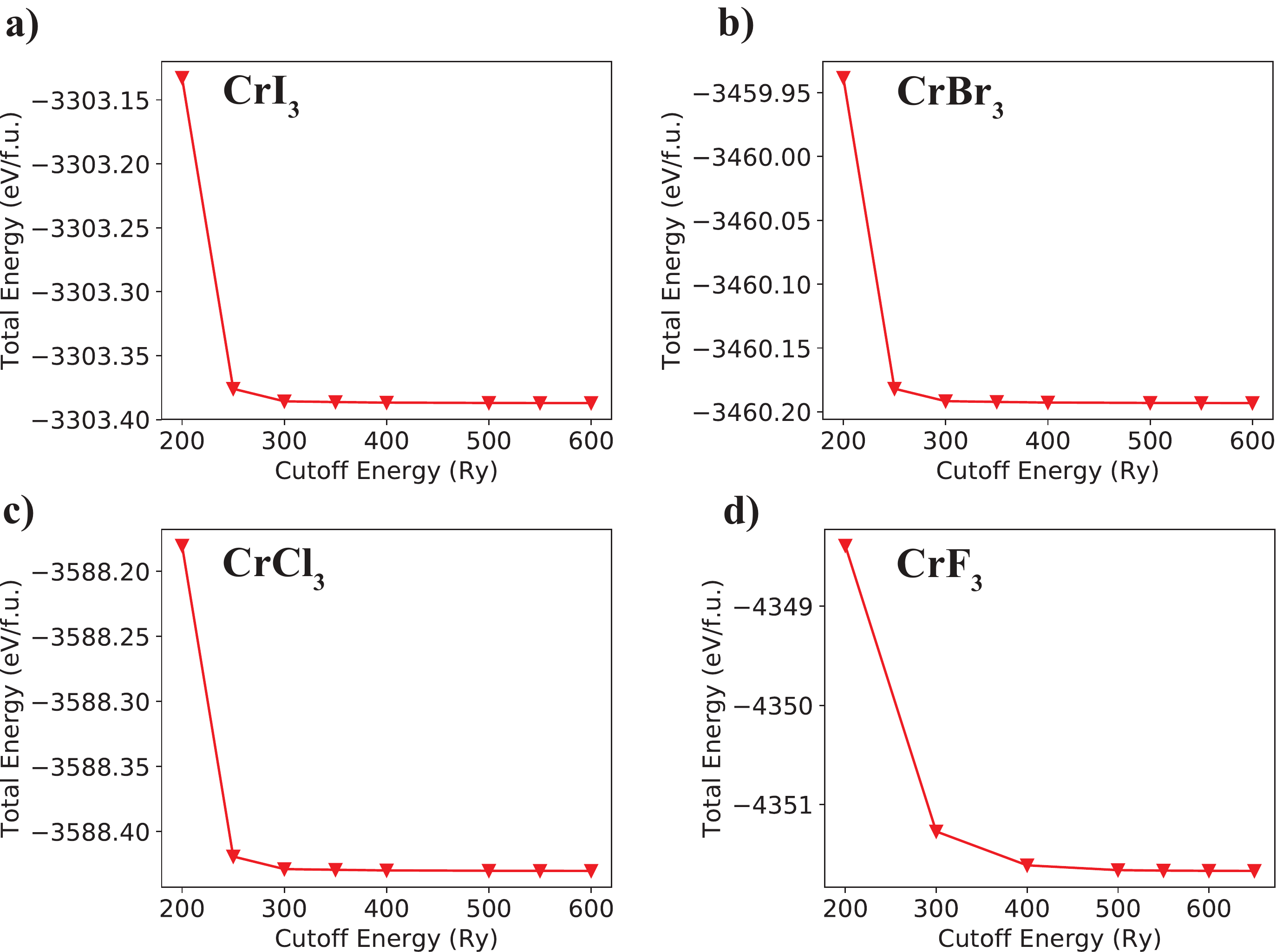}
\caption{The total energy per formula unit of the unit cell (8 atoms) of 2D a) CrI$_3$, b) CrBr$_3$, c) CrCl$_3$, and d) CrF$_3$ as a function of plane wave cutoff energy
for the norm-conserving pseudopotentials calculated with density functional theory (DFT) using the Perdew-Burke-Ernzerhof (PBE) functional at a single K-point.
The results show a converged cutoff energy of 300 Ry for CrI$_3$, CrBr$_3$ and CrCl$_3$ and 600 Ry for CrF$_3$.}
\label{cutoff}
\end{center}
\end{figure}

\begin{figure}
\begin{center}
\includegraphics[width=14cm]{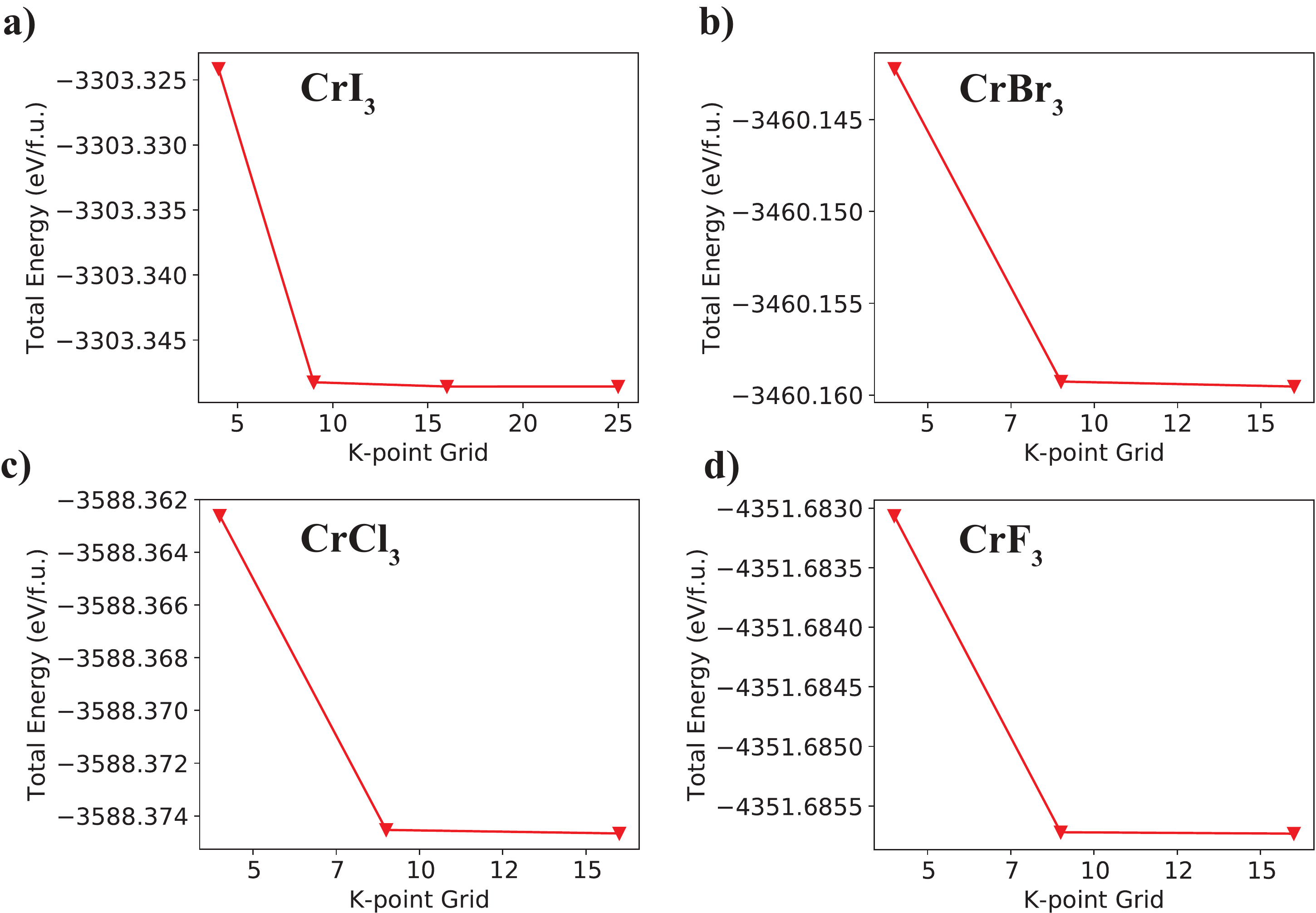}
\caption{The total energy per formula unit of the unit cell (8 atoms) of 2D a) CrI$_3$, b) CrBr$_3$, c) CrCl$_3$, and d) CrF$_3$ as a function of K-point grid
for the norm-conserving pseudopotentials calculated with DFT (PBE) at the converged cutoff energy (see Fig. S1).
The results show a converged k-point grid of 3x3x1 (9) for all monolayers. The number of k-points was scaled appropriately to obtain the converged grid depending on the supercell size and shape for all DFT and diffusion Monte Carlo (DMC) calculations.}
\label{kpoint}
\end{center}
\end{figure}

\begin{table}[]
\caption{
Tabulated results for the DMC timestep convergence of the unit cell (8 atoms) of 2D CrI$_3$, CrBr$_3$, CrCl$_3$, and CrF$_3$. The acceptance ratio of 0.99 indicates that 0.01 Ha$^{-1}$ is an appropriate timestep to use for all subsequent DMC simulations.   }
\begin{tabular}{l|l|l|l}
\hline
\hline
CrI3             &                      &            &                  \\
\hline
\hline
Timestep (Ha$^{-1}$) & DMC Total Energy (Ha) & Error (Ha) & Acceptance Ratio \\
\hline
0.05                                 & -242.597                                  & 0.001                          & 0.96                                 \\
0.02                                 & -242.543                                  & 0.001                          & 0.99                                 \\

0.01                                 & -242.527                                  & 0.001                          & 0.99                                 \\
0.005                                & -242.529                                  & 0.002                          & 1.00                                 \\
0.001                                & -242.531                                  & 0.002                          & 1.00                                 \\
\hline
\hline
CrBr3             &                      &            &                  \\
\hline
\hline
Timestep (Ha$^{-1}$) & DMC Total Energy (Ha) & Error (Ha) & Acceptance Ratio \\
\hline
0.05                                 & -255.368                                  & 0.311                          & 0.95                                 \\
0.02                                 & -254.194                                  & 0.002                          & 0.98                                 \\

0.01                                 & -254.181                                  & 0.001                          & 0.99                                 \\
0.005                                & -254.183                                  & 0.002                          & 1.00                                 \\
0.001                                & -254.185                                  & 0.002                          & 1.00                                 \\
\hline
\hline
CrCl3             &                      &            &                  \\
\hline
\hline
Timestep (Ha$^{-1}$) & DMC Total Energy (Ha) & Error (Ha) & Acceptance Ratio \\
\hline
0.05                                 & -263.943                                  & 0.001                          & 0.95                                 \\
0.02                                 & -263.878                                  & 0.002                          & 0.98                                 \\
 
0.01                                 & -263.869                                  & 0.001                          & 0.99                                 \\
0.005                                & -263.866                                  & 0.001                          & 1.00                                 \\
0.001                                & -263.867                                  & 0.003                          & 1.00                                 \\
\hline
\hline
CrF3             &                      &            &                  \\
\hline
\hline
Timestep (Ha$^{-1}$) & DMC Total Energy (Ha) & Error (Ha) & Acceptance Ratio \\
\hline
0.05                                 & -319.919                                  & 0.001                          & 0.92                                 \\
0.02                                 & -319.842                                  & 0.001                          & 0.97                                 \\

0.01                                 & -319.834                                  & 0.001                          & 0.99                                 \\
0.005                                & -319.836                                  & 0.002                          & 0.99                                 \\
0.001                                & -319.820                                  & 0.003                          & 1.00                                
\end{tabular}
\end{table}

\begin{figure}
\begin{center}
\includegraphics[width=12cm]{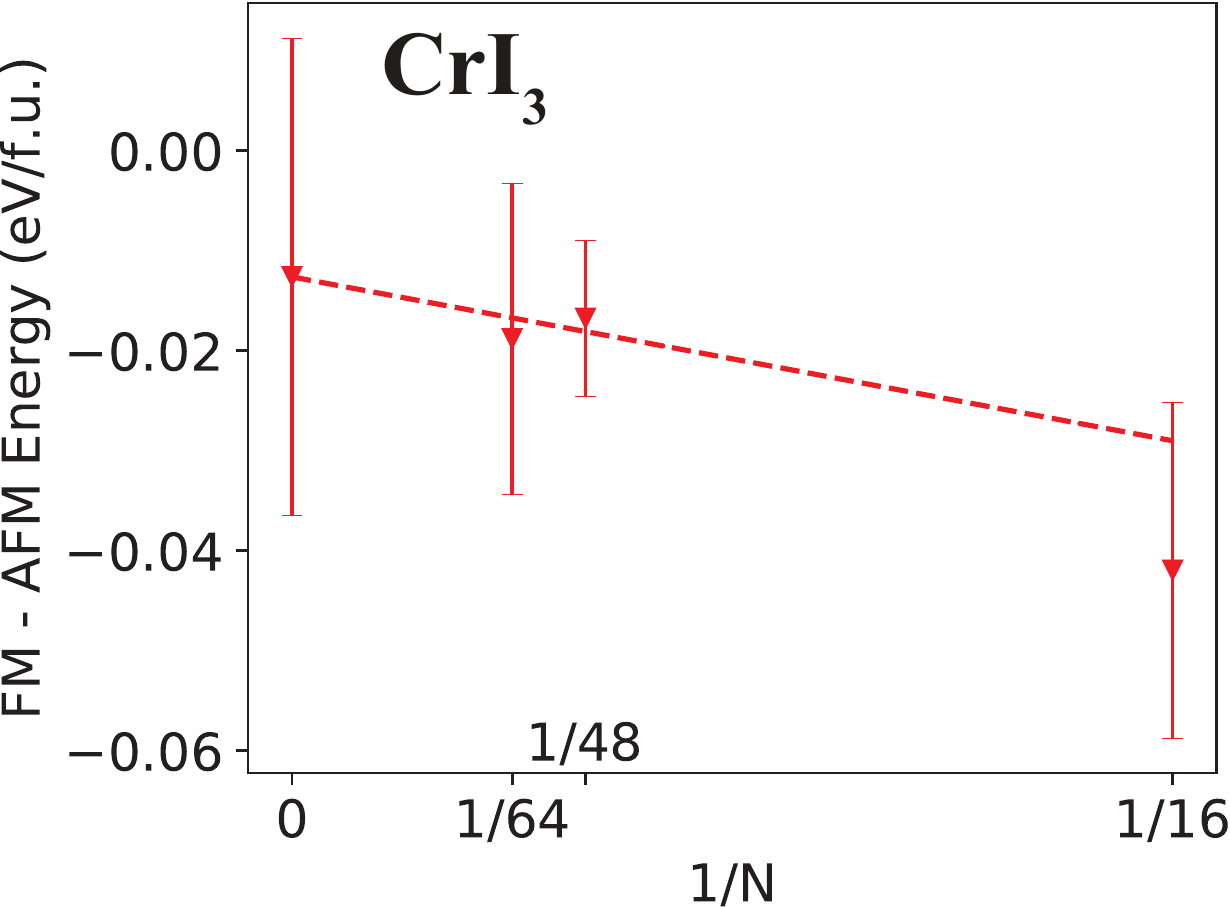}
\caption{The ferromagnetic - antiferromagnetic energy per formula unit of the CrI$_3$ as a function of supercell size (N = number of atoms) to demonstrate finite size convergence. These results indicate that a 48 atom cell is sufficient to determine $J$, and therefore a 48 atom cell was used to calculate $J$ for CrI$_3$ and CrBr$_3$. The error bars represent the standard error about the mean value.  }
\label{fs}
\end{center}
\end{figure}

\begin{table}[]
\caption{
Tabulated results for the structural parameters of CrI$_3$, CrBr$_3$, CrCl$_3$, and CrF$_3$ relaxed with each DFT functional using the Vienna Ab initio Simulation code (presented in Table 2 in the main text). $\theta_1$ and $\theta_2$ are defined in Fig. 1 in the main text.   }
\begin{adjustbox}{width=10cm,center}
\begin{tabular}{l|l|l|l|l|l}
\hline
\hline
CrI$_3$        &        &         &     &       &         \\
\hline
\hline
Functional  & $a$ (\AA)       & $d_{Cr-X}$ (\AA)     & $d_{Cr-Cr}$ (\AA)  & $\theta_1$  & $\theta_2$   \\
\hline
vdW-DF-OptB88 & 6.90                        & 2.73                         & 3.98                      & 90.35                       & 175.16                       \\
PBE         & 7.00                        & 2.74                         & 4.04                      & 90.63                       & 173.24                       \\
PBE+U       & 7.05                        & 2.76                         & 4.07                      & 90.51                       & 173.43                       \\
LDA         & 6.69                        & 2.66                         & 3.86                      & 90.30                       & 175.55                       \\
LDA+U       & 6.73                        & 2.68                         & 3.89                      & 90.15                       & 175.77                       \\
SCAN        & 6.96                        & 2.73                         & 4.02                      & 90.61                       & 174.07                       \\
SCAN+U      & 6.99                        & 2.75                         & 4.03                      & 90.49                       & 174.40                       \\
r2SCAN      & 6.98                        & 2.74                         & 4.03                      & 90.57                       & 174.08                       \\
r2SCAN+U    & 7.02                        & 2.76                         & 4.05                      & 90.48                       & 174.33                       \\
DMC\cite{staros}         & 6.87(3) & 2.722(8) &     & 90.4(2) & 175.4(3) \\
\hline
\hline
CrBr$_3$       &        &         &      &         &          \\
\hline
\hline
Functional  & $a$ (\AA)       & $d_{Cr-X}$ (\AA)     & $d_{Cr-Cr}$ (\AA)  & $\theta_1$  & $\theta_2$   \\
\hline
vdW-DF-OptB88 & 6.34                        & 2.51                         & 3.66                      & 90.41                       & 175.15                       \\
PBE         & 6.44                        & 2.52                         & 3.72                      & 90.68                       & 173.51                       \\
PBE+U       & 6.48                        & 2.54                         & 3.74                      & 90.58                       & 173.63                       \\
LDA         & 6.17                        & 2.45                         & 3.56                      & 90.35                       & 175.57                       \\
LDA+U       & 6.21                        & 2.47                         & 3.58                      & 90.22                       & 175.71                       \\
SCAN        & 6.39                        & 2.50                         & 3.69                      & 90.39                       & 174.07                       \\
SCAN+U      & 6.42                        & 2.52                         & 3.71                      & 90.57                       & 174.24                       \\
r2SCAN      & 6.39                        & 2.51                         & 3.69                      & 90.61                       & 174.34                       \\
r2SCAN+U    & 6.42                        & 2.53                         & 3.71                      & 90.53                       & 174.48                       \\
\hline
\hline
CrCl$_3$       &        &         &      &       &       \\
\hline
\hline
Functional  & $a$ (\AA)       & $d_{Cr-X}$ (\AA)     & $d_{Cr-Cr}$ (\AA)  & $\theta_1$  & $\theta_2$   \\
\hline
vdW-DF-OptB88 & 5.98                        & 2.35                         & 3.45                      & 90.87                       & 174.75                       \\
PBE         & 6.06                        & 2.36                         & 3.50                      & 91.10                       & 173.24                       \\
PBE+U       & 6.10                        & 2.38                         & 3.52                      & 91.03                       & 173.34                       \\
LDA         & 5.82                        & 2.30                         & 3.36                      & 90.78                       & 175.10                       \\
LDA+U       & 5.86                        & 2.31                         & 3.38                      & 90.69                       & 175.27                       \\
SCAN        & 5.99                        & 2.35                         & 3.46                      & 90.89                       & 174.03                       \\
SCAN+U      & 6.02                        & 2.36                         & 3.47                      & 90.76                       & 174.18                       \\
r2SCAN      & 6.00                        & 2.35                         & 3.47                      & 90.93                       & 174.11                       \\
r2SCAN+U    & 6.04                        & 2.37                         & 3.49                      & 90.93                       & 174.20                       \\
Exp &       5.942(3)\cite{doi:10.1063/1.1725428}                  &                         &                       &                       &                        \\
      \hline
      \hline
CrF$_3$        &        &         &      &        &         \\
\hline 
\hline
Functional  & $a$ (\AA)       & $d_{Cr-X}$ (\AA)     & $d_{Cr-Cr}$ (\AA)  & $\theta_1$  & $\theta_2$   \\
\hline
vdW-DF-OptB88 & 5.14                        & 1.94                         & 2.97                      & 94.61                       & 172.13                       \\
PBE         & 5.19                        & 1.94                         & 3.00                      & 94.88                       & 171.27                       \\
PBE+U       & 5.23                        & 1.95                         & 3.02                      & 94.90                       & 171.10                       \\
LDA         & 5.04                        & 1.90                         & 2.91                      & 94.83                       & 172.44                       \\
LDA+U       & 5.07                        & 1.91                         & 2.93                      & 94.84                       & 172.32                       \\
SCAN        & 5.13                        & 1.92                         & 2.96                      & 94.81                       & 171.55                       \\
SCAN+U      & 5.17                        & 1.93                         & 2.99                      & 94.99                       & 171.22                       \\
r2SCAN      & 5.13                        & 1.92                         & 2.96                      & 94.79                       & 171.69                       \\
r2SCAN+U    & 5.17                        & 1.94                         & 2.99                      & 94.83                       & 171.40                      
\end{tabular}
\end{adjustbox}
\end{table}

\begin{figure}
\begin{center}
\includegraphics[width=16cm]{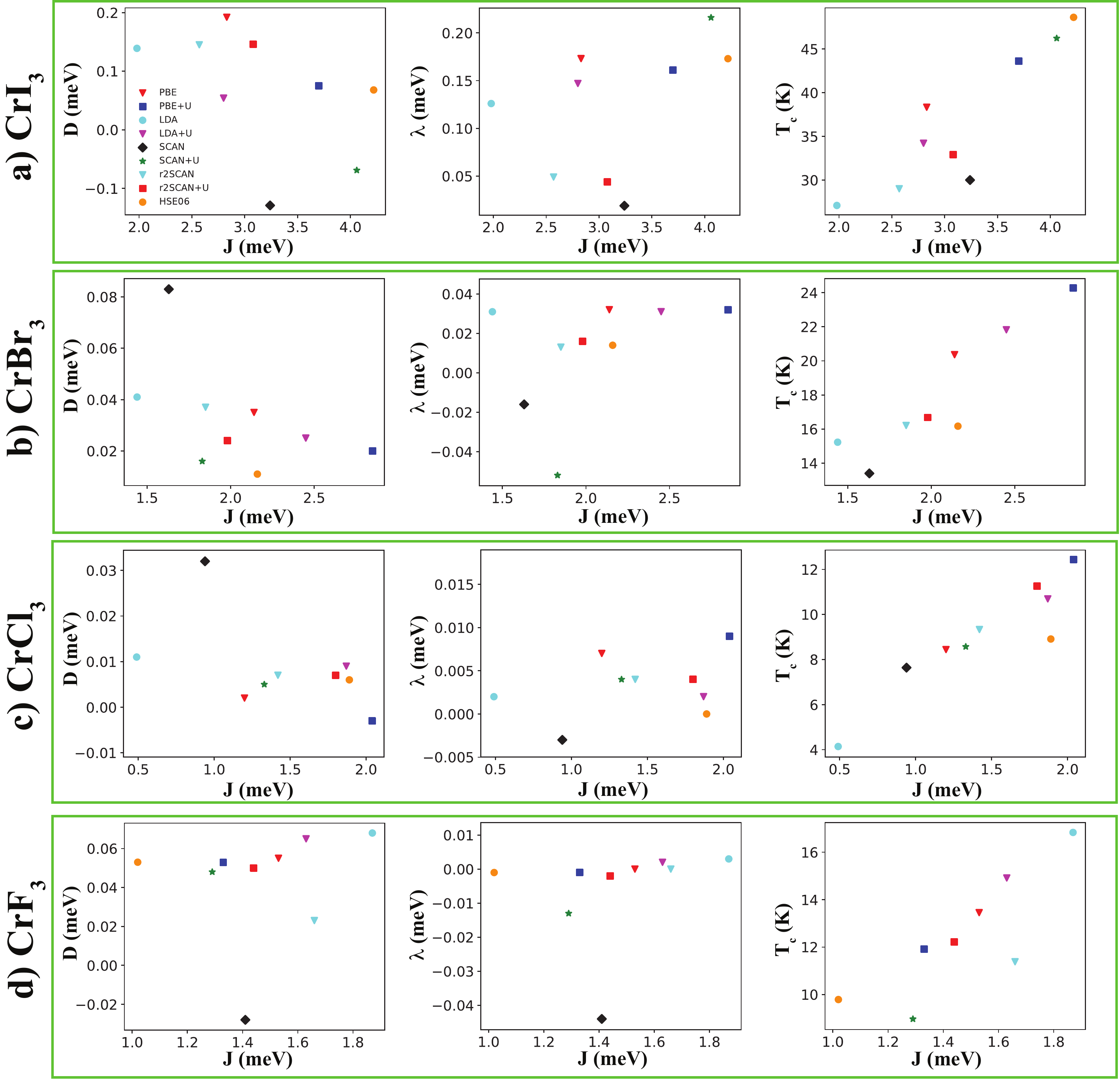}
\caption{A scatter plot of the DFT benchmarking data presented in Table 1 (fixed vdW-DF-OptB88 geometry) for a) CrI$_3$, b) CrBr$_3$, c) CrCl$_3$ and c) CrF$_3$.  }
\label{vasp}
\end{center}
\end{figure}

\begin{figure}
\begin{center}
\includegraphics[width=16cm]{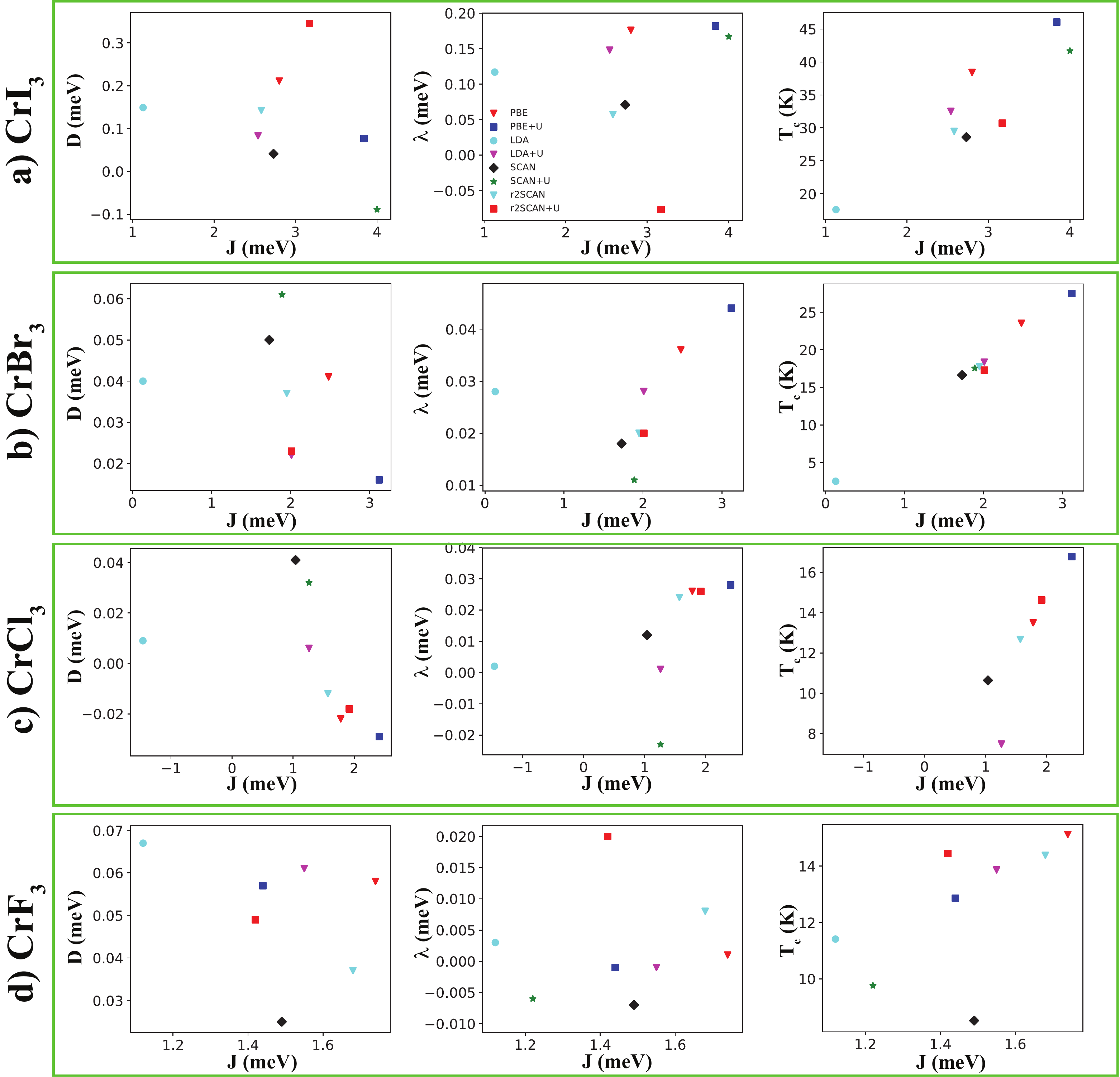}
\caption{A scatter plot of the DFT benchmarking data presented in Table 2 (relaxed geometry) for a) CrI$_3$, b) CrBr$_3$, c) CrCl$_3$ and c) CrF$_3$. }
\label{vasp2}
\end{center}
\end{figure}

\begin{table}[]
\caption{
Tabulated results for $J$ and T$_c$ from literature in comparison to the DMC values from our own work. The T$_c$ values are determined by the Torelli and Olsen model unless noted by *, which indicates these were calculated with spin wave theory.  }
\begin{tabular}{l|l|l}
\hline
\hline
CrI3     &   &                         \\
\hline
\hline
Method   & $J$ & Tc                      \\
\hline
\cite{Lado_2017} PBE+2    & 2.20                  &  33* \\
\hline
\cite{Torelli_2018}  PBE      &  &  32\\

\cite{Torelli_2018}  PBE+1    &  &   37\\
\cite{Torelli_2018}  PBE+2    & 3.25                  &   42\\
\cite{Torelli_2018}  PBE+3    &  &  47\\
\hline
\cite{PhysRevLett.127.166402}BSE(LDA) & 2.40                  &                         \\
\hline
\cite{Kumar_Gudelli_2019}PBE+2    & 3.07                  &   66*\\
\hline
\cite{kitaev}LDA+0.5  & 2.29                  &                         \\
\hline
\cite{olsen-data}PBE      & 1.94                  &   28\\
\hline
\cite{PhysRevB.98.144411}PBE      & 2.70                  &                         \\
\hline
\cite{C5TC02840J}HSE06    & 3.32                  &                         \\

\cite{C5TC02840J}PBE      & 2.71                  &                         \\
\hline
\cite{Pizzochero_2020}MRCI     & 2.88                  &                         \\
\hline
DMC &  2.49(1.16)     &                         \\
\hline
\hline
CrBr3    &   &                         \\
\hline
\hline
Method   & $J$ & Tc                      \\
\hline
\cite{olsen-data}  PBE      & 1.84                  &   19\\
\hline
\cite{PhysRevB.98.144411}PBE      & 2.40                  &                         \\
\hline
\cite{C5TC02840J}HSE06    & 2.42                  &                         \\

\cite{C5TC02840J}PBE      & 2.45                  &                         \\
\hline
DMC  & 1.30(1.00)      &                         \\
\hline
\hline
CrCl3    &   &                         \\
\hline
\hline
Method   & $J$ & Tc                      \\
\hline
\cite{olsen-data}PBE      & 1.29                  & 9.2 \\
\hline
\cite{PhysRevB.98.144411}PBE      & 1.70                  &                         \\
\hline
\cite{C5TC02840J}HSE06    & 1.87                  &                         \\
\cite{C5TC02840J}PBE      & 1.79                  &                         \\
\hline
\hline
CrF3     &   &                         \\
\hline
\hline
Method   & $J$ & Tc                      \\
\hline
\cite{C5TC02840J}HSE06    & 0.87                  &                         \\
\cite{C5TC02840J}PBE      & 1.79                  &       \\
\hline
\hline
\end{tabular}
\end{table}

\begin{figure}
\caption{DMC
calculated total energies of a 16-atom supercell (normalized per formula unit (f.u.)) of the ferromagnetic orientation of 2D a) CrCl$_3$ and b) CrF$_3$ calculated as a function of the U parameter used to variationally determine the optimal trial wave function. For
convenience of presentation, the DMC energies are shifted by the lowest DMC energy obtained at the appropriate U value (U = 1 eV for CrCl$_3$ and U = 4 eV for CrF$_3$). The error bars represent the standard error about the mean value. }
\begin{center}
\includegraphics[width=8cm]{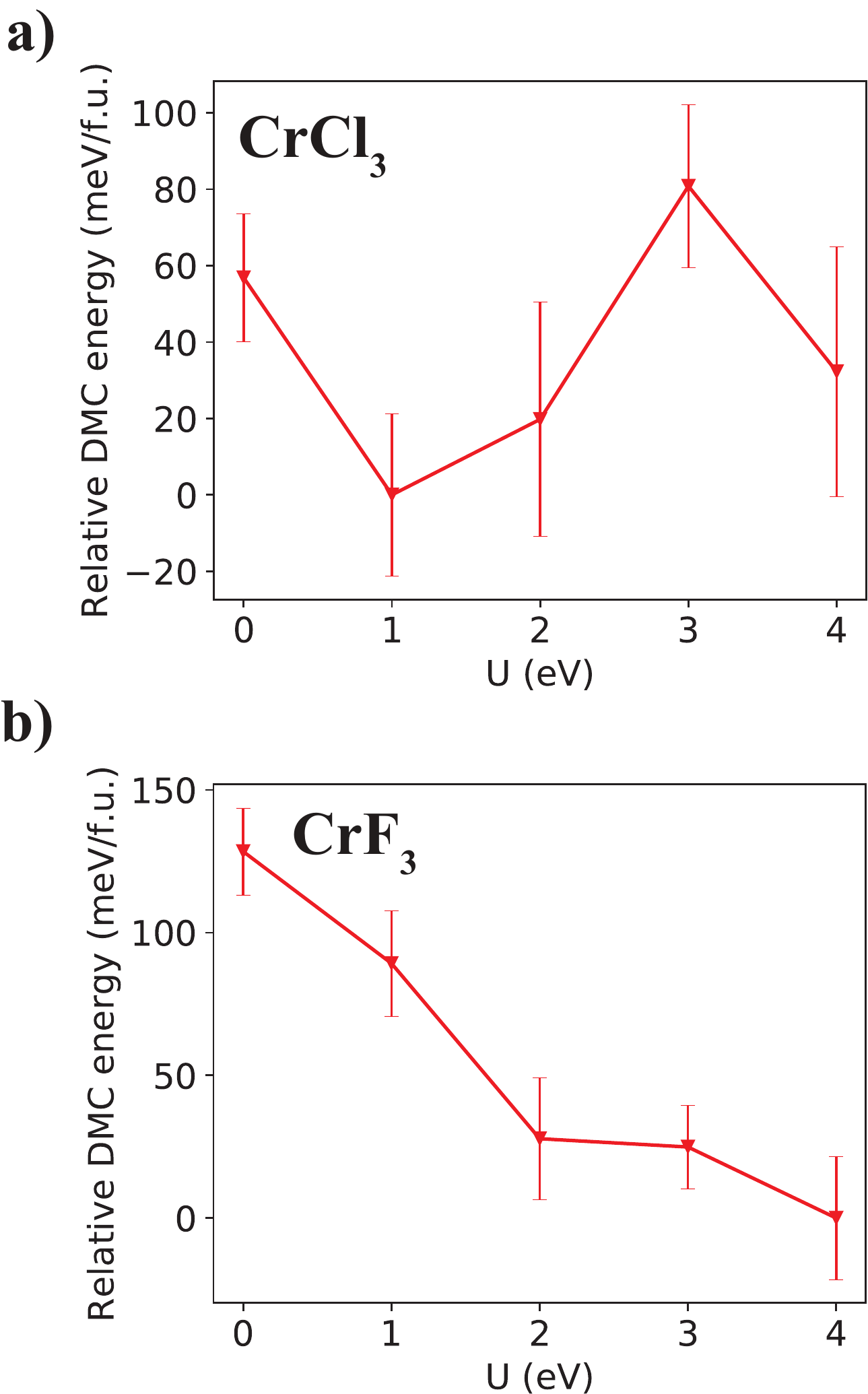}
\label{u-tune-si}
\end{center}
\end{figure}

\begin{figure}
\begin{center}
\includegraphics[width=15cm]{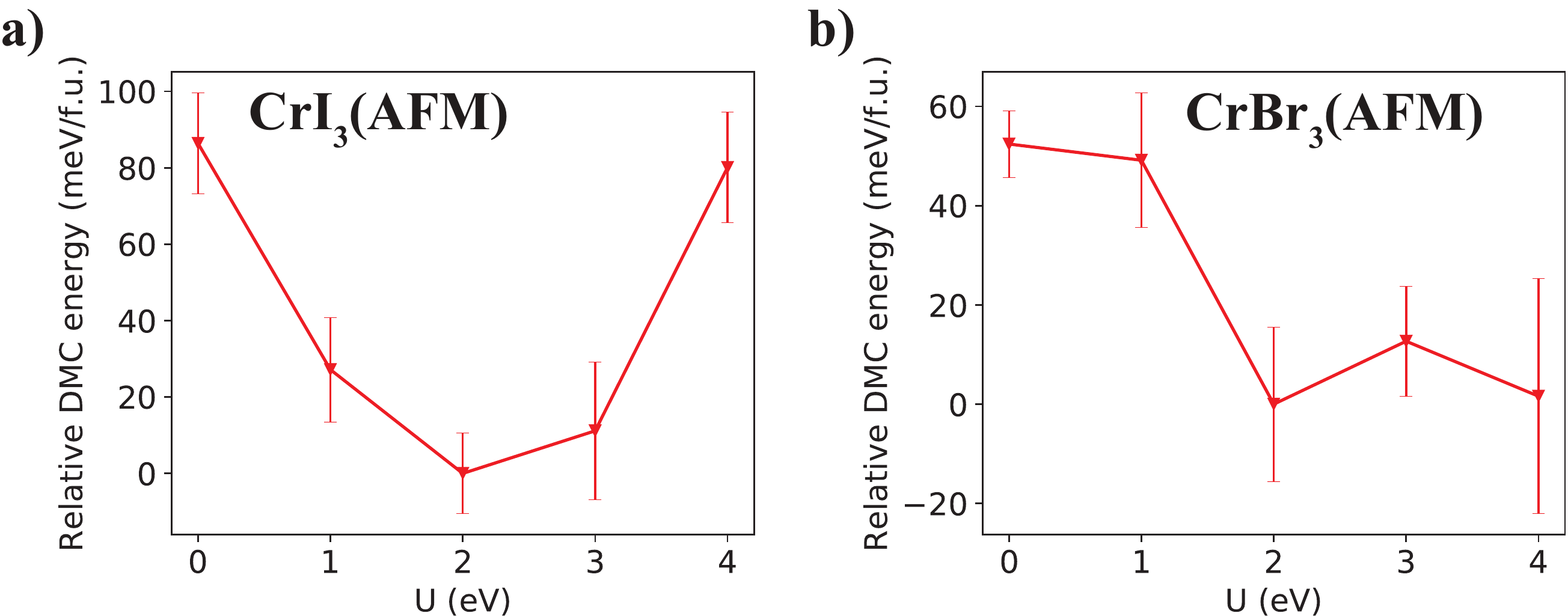}
\caption{DMC
calculated total energies and subsequent standard error about the mean (indicated by error bars) of a 16-atom supercell (normalized per formula unit (f.u.)) of the antiferromagnetic orientation of 2D a) CrI$_3$ and b) CrBr$_3$ calculated as a function of the U parameter used to variationally determine the optimal trial wave function. For
convenience of presentation, the DMC energies are shifted by the lowest DMC energy obtained at the appropriate U value (U = 2 eV for CrI$_3$ and CrBr$_3$).   }
\label{u-tune}
\end{center}
\end{figure}

\begin{table}[]
\caption{
Tabulated results that illustrate the range of T$_c$ calculated from a fixed value of $J$ estimated from DMC and various values for magnetic anisotropy calculated with different DFT functionals.   }
\begin{adjustbox}{width=16cm,center}
\begin{tabular}{l|lllllllll}
\hline
CrI$_3$         & \textbf{Anisotropy} &       &    &      &     &       &       &         &      \\
\hline
$J$ = 2.49 meV & PBE        & PBE+U & LDA & LDA+U & SCAN & SCAN+U & r2SCAN & r2SCAN+U & HSE06 \\
\hline
$D$ (meV)      & 0.19                           & 0.08                      & 0.14                    & 0.05                      & -0.13                    & -0.07                      & 0.14                       & 0.15                         & 0.07                      \\
$\lambda$ (meV) & 0.17                           & 0.16                      & 0.13                    & 0.15                      & 0.02                     & 0.22                       & 0.05                       & 0.04                         & 0.17                      \\
$\Delta$ (meV)  & 1.16                           & 0.87                      & 0.85                    & 0.77                      & -0.17                    & 0.84                       & 0.51                       & 0.49                         & 0.91                      \\
\hline
\textbf{T$_c$ (K)}       & 34.71(11.57)                          & 32.34(10.78)                     & 32.07(10.69)                   & 31.31(10.43)                     & -    & 31.98(10.66)                      & 28.29(9.42)                      & 28.02(9.33)                        & 32.67(10.89)                     \\
\hline
\hline
CrBr$_3$         & \textbf{Anisotropy} &       &    &      &     &       &       &         &      \\
\hline
$J$ = 1.30 meV & PBE        & PBE+U & LDA & LDA+U & SCAN & SCAN+U & r2SCAN & r2SCAN+U & HSE06 \\
\hline
$D$ (meV)      & 0.04                           & 0.02                      & 0.04                    & 0.03                      & 0.08                     & 0.02                       & 0.04                       & 0.02                         & 0.01                      \\
$\lambda$ (meV) & 0.03                           & 0.03                      & 0.03                    & 0.03                      & -0.02                    & -0.05                      & 0.01                       & 0.02                         & 0.01                      \\
$\Delta$ (meV)  & 0.21                           & 0.18                      & 0.22                    & 0.19                      & 0.09                     & -0.20                      & 0.13                       & 0.12                         & 0.08                      \\
\hline
\textbf{T$_c$ (K)}        & 13.95(7.44)                          & 13.43(7.17)                     & 14.07(7.51)                   & 13.54(7.23)                     & 11.30(6.03)                    & -      & 12.41(6.62)                      & 12.13(6.47)                        & 11.02(5.88)                 \\
\hline
\end{tabular}
\end{adjustbox}
\end{table}

\begin{figure}
\caption{Total radially averaged charge density calculated with DMC and PBE+U (U = [0, 1, 2, 3, 4] eV) of a) Cr and b) I for 2D CrI$_3$ and c) Cr and d) Br for 2D CrBr$_3$. The error bars represent the standard error about the mean value. }
\begin{center}
\includegraphics[width=15cm]{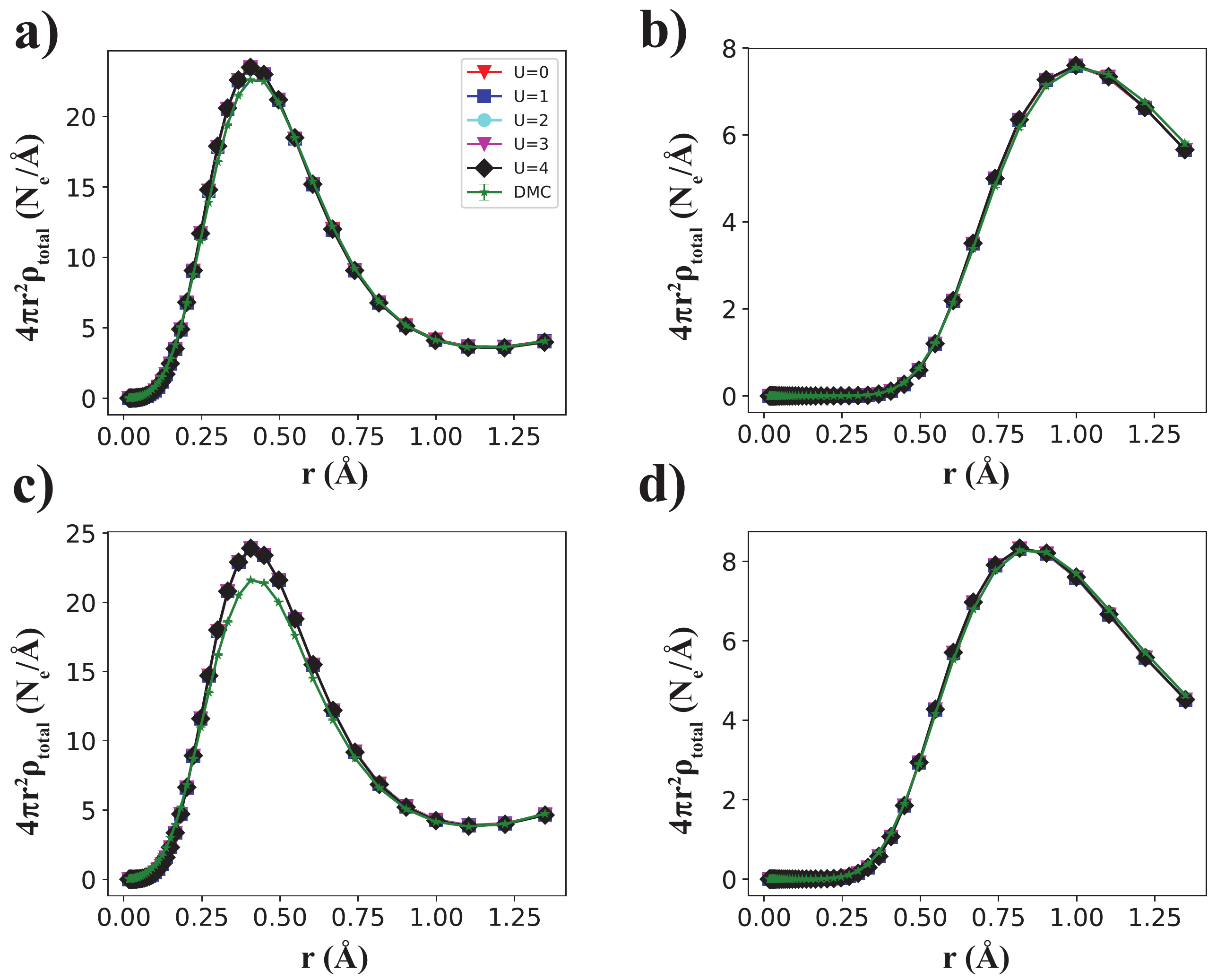}
\label{dens}
\end{center}
\end{figure}

The diffusion Monte Carlo (DMC) simulations performed to obtain an accurate estimate of magnetic exchange ($J$) were conducted on 8 nodes of our local cluster. Each node is connected via OmniPath networking and contains 2 Intel 20-Core Xeon Gold 6138 CPUs clocked at 2GHz and 384 GB of 2666 MHz DDR4 memory. QMCPack version 3.9.0 is compiled using OpenMPI version 3.0.1 and Intel Fortran, C and C++ (version 19.0.4.243) compiler.

\newpage

\bibliography{si}% Produces the bibliography via BibTeX.